# Real GDP per capita since 1870

**Ivan Kitov**, Institute for the Geospheres' Dynamics, RAS
**Oleg Kitov**, University of Oxford

**Abstract**
The growth rate of real GDP per capita in the biggest OECD countries is represented as a sum of two components – a steadily decreasing trend and fluctuations related to the change in some specific age population. The long term trend in the growth rate is modelled by an inverse function of real GDP per capita with a constant numerator. This numerator is equivalent to a constant annual increment of real GDP per capita. For the most advanced economies, the GDP estimates between 1950 and 2007 have shown very weak and statistically insignificant linear trends (both positive and negative) in the annual increment. The fluctuations around relevant mean increments are characterized by practically normal distribution. For many countries, there exist historical estimates of real GDP since 1870. These estimates extend the time span of our analysis together with a few new estimates from 2008 to 2011. There are severe structural breaks in the corresponding time series between 1940 and 1950, with the slope of linear regression increasing by a factor of 4.0 (Switzerland) to 22.1 (Spain). Therefore, the GDP estimates before 1940 and after 1950 have been analysed separately. All findings of the original study are validated by the newly available data. The most important is that all slopes (except that for Australia after 1950) of the regression lines obtained for the annual increments of real GDP per capita are small and statistically insignificant, i.e. one cannot reject the null hypothesis of a zero slope and thus constant increment. Hence the growth in real GDP per capita is a linear one since 1870 with a break in slope between 1940 and 1950.

**Key words:** GDP, model, economic growth, inertia, trend, OECD

**Introduction**

There are many problems associated with the theory of long-term economic growth. The study of Galor (2005) describes the evolution of income per capita since the epoch of Malthusian stagnation and discusses the process, which induced the transition to the current sustained economic growth in developed countries. The primary aim of Galor's study, as well as economics as a whole, is to find a unifying theory accommodating various periods of growth based on solid micro foundations. This is a fundamental approach which should be also supported by observations at macro level.

In this study, we validate our original model (Kitov, 2008) describing the evolution of real GDP per capita in developed countries between 1950 and 2007. An extended historical dataset is used as provided by Angus Maddison (2004) and the Groningen Growth and Development Centre. In this dataset, there are some continuous time series of real GDP per capita since 1820. We have selected the year of 1870 as a starting point. All studied countries are characterized by continuous real GDP time series since this year. The Maddison historical data are available only to 2008. Therefore, we use the Conference Board (2012) Total Economy Database (TED) for the period between 1950 and 2011.

We do not use any sophisticated technique of signal extraction, as proposed by Pedregal (2005), who explored two linear trend models with a nonlinear forecast function. Under our framework, the long-term forecast is not limited in time but is based on a constant annual increment of real GDP per capita, which we call the inertial growth due to its similarity to inertia in classical mechanics. In other words, a developed economy will be growing with a constant annual increment of real GDP per capita when no external forces are applied. Similarly, an object will continue moving at its current velocity (distance increment per unit time) until external forces are applied. Obviously, our model does not allow an exponential growth path, unlike that presumed in the trend extracting procedures developed by Pollock (2007) or in the original Solow (1956) and Swan (1956) growth model.

In this article, we show that annual increment of real GDP per capita has no statistically reliable linear trend in thirteen developed economies (from Australia to the US) since 1870. There are two distinct periods where real GDP grows linearly with time (i.e. the annual increment is constant) – from 1870 to 1940 and after 1950. There is a break in all involved time series between 1940 and 1950 which is likely related to the change in measuring units and procedures. Before 1940, the GDP estimates are based on data not fully relevant to the canonical definition of Gross Domestic Product as introduced and developed in the 1930s and 1940s. Therefore, the corresponding GDP values are mainly reconstructed not measured. Nevertheless, they also demonstrate linear growth paths of GDP per capita.

The remainder of the article consists of three Sections and Conclusion. Section 1 describes the concept of real GDP evolution and introduces several models predicting the path of real economic growth. Section 2 revisits the original results obtained from 1950 to 2007 and extends the data set to 2011. All original regressions are recalculated and statistically assessed for the extended period. In Section 3, we process the data between 1870 and 1940 using the same approach.

### 1. The model

The real problem with the description of economic processes is the inability of other sciences, including physics, to outperform economics despite sporadic claims (Bouchard, 2008). Without a valid quantitative theory of macroeconomic processes (in sense that *ex-ante* predictions are consistent with *ex-post* observations) the current hopeless situation may last for a long time. To be valid and useful, any scientific theory must fit observations and predict new effects or future evolution. Unfortunately, the current economic paradigm repudiates, without any formal or empirical proof, the possibility to develop a deterministic economic theory. Our main aim is to show that such theories can indeed be formulated and should not be overlooked. In our attempt to do so, we will begin by presenting a deterministic model for real GDP per capita.

Let us start by introducing a new concept describing the evolution of real Gross Domestic Product. Our central claim is quite straightforward – the growth rate, *g(t)*, of real GDP per capita, *G(t)*, is driven by the attained level of real GDP per capita and the change in a specific age population, $N_s$. According to this model, the growth rate of real GDP (for the sake of brevity we often omit "per capita") in developed countries is primarily characterized by an annual increment, *A*, which does not change over time; *A = const*. All fluctuations around this constant increment can be explained by the change in the number of people of the country-specific age:

$$g(t) = dlnG(t)/dt = A/G(t) + 0.5dlnN_s(t)/dt \qquad (1)$$

Equation (1) is a quantitative model that has been constructed empirically and proved statistically by tests for cointegration (Kitov, Kitov, and Dolinskaya, 2009). Notice that unlike in the mainstream economics, no assumptions were made on the behaviour of economic agents and no theoretical models were formulated at the initial stage of our empirical study. Instead we have attempted to find a model that would best fit observations. The next subsection will be devoted to discussion of the actual model fitting.

We use the relative growth rate, as represented by *dG(t)/G(t)=dlnG(t)*. In order to better understand the processes defining real growth let us decompose the model and consider each term individually. We will for now assume that the second term in (1), i.e. the rate of change in the specific age population, is zero. Accordingly, there is no external force acting on the GDP growth rate and the system is in the state of stationary or inertial (in terms of constant increment per unit time) growth. Later on we provide a simple analogue from mechanics, which clarifies why we prefer to call *A/G(t)* "the inertial growth".

When the population age pyramid is fixed ($dN_s \equiv 0$), real GDP grows as a linear function of time:

$$g(t) = dlnG(t)/dt (given\ dN_s(t) = 0) = A/G(t)$$
$$G(t) = At + C \qquad (2)$$

where *G(t)* is completely equivalent to the inertial growth, $G_i(t)$, i.e. the first component of the overall growth as defined by (1). Relationship (2) defines the linear trajectory of the GDP per capita, where $C=G_i(t_0)=G(t_0)$ and $t_0$ is the starting time. In the regime of inertial growth, the real GDP per capita increases by the constant value $A$ per time unit. Relationship (3) is equivalent to (2), but holds for the inertial part of the total growth:

$$G_i(t) = G_i(t_0) + At \qquad (3)$$

The relative rate of growth along the inertial linear growth trend, $g_i(t)$, is the reciprocal function of $G_i$ or, equivalently, $G$:

$$g_i(t) = dlnG_i/dt = A/G_i = A/G(t) \qquad (4)$$

Relationship (4) implies that the rate of GDP growth will be asymptotically approaching zero, but the annual increment $A$ will be constant. Moreover, the absolute rate of GDP growth is constant and is equal to $A$ [$/y]. This constant annual increment thus defines the constant "speed" of economic growth in a one-to-one analogy with Newton's first law. Hence, one can consider the property of constant speed of real economic growth as "inertia of economic growth" or simply "inertia". Then the growth, which is observed without the change in the specific age population, can be called the "inertial growth".

A textbook analogy of inertia at work from mechanics is rotation of a mass on a rope. Rotation around the centre is accompanied by the change in the direction of motion and is driven by the tension force in the rope. If suddenly the rope is broken the mass follows up linear progressive motion at a constant speed along the line defined by the velocity vector at the moment when the rope was broken. In other words, the mass continues inertial motion with inertia being the property allowing for constant speed and direction. This works only when there is no net force to change the speed and direction. However, in order to retain constant speed and direction in real world (e.g. a plane flight) one needs to supply nonzero forces to compensate all traction forces.

In physics, inertia is the most fundamental property. In economics, it should also be a fundamental property, taking into account the difference between ideal theoretical equilibrium of space/time and the stationary real behaviour of the society. Mechanical inertia implies that no change in motion occurs in the absence of net exogenous force and without change in internal energy. (As mentioned above, in real case the net force is zero but one should apply extra forces in order to overcome the net traction force and to keep the body moving at a constant speed.) For a society, the net force applied by all economic agents is not zero but counteracts all dissipation processes and creates goods and services in excess of the previous level. The economy does grow with time and its "internal energy" as expressed in monetary units does increase at a constant speed.

We do not consider the economy as a stone flying through space at a constant speed. The economy is a complex system with all internal forces providing constant speed of growth. The stone has no internal forces, which are able to change its speed. In reality, the space is full of dust and electromagnetic fields which can change the speed. The economy has more traction forces, bumps and barriers. That's why the speed of inertial growth differs between developed countries as we have confirmed empirically. Moreover, not all economies are organized in a way that results in the optimal stationary behaviour and the highest speed of economic growth.

Let us now consider the second growth component – the relative rate of change in the number of "*s*"-year-olds. As a rule, in Western Europe the integral change in the specific age population during the last 60 years is negligible, and thus, the cumulative input of the population component is close to zero. In the USA, the overall increase in the specific age population is responsible for about 20% of the total growth in real GDP per capita since 1960. In (1), the term *$0.5 d\ln N_s(t)/dt$* is the halved rate of growth in the number of *s*-year-olds at time *t*. The factor of 1/2 is common for developed countries. The only exception we have found so far and report later in this Section is Japan, where this factor is 2/3 as obtained from the rate of growth.

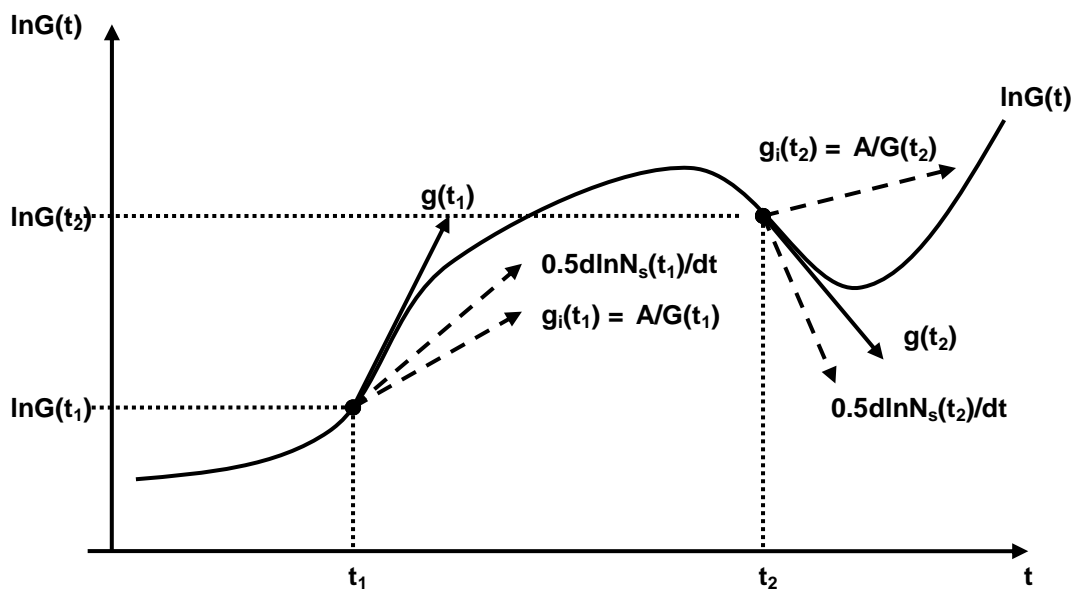

**Figure 1. Illustration of the growth model for real GDP per capita**

Figure 1 depicts an arbitrary GDP evolution curve, *lnG(t)*, which exhibits episodes of rapid growth ($t_1$) and recession ($t_2$). It is easier to illustrate the performance of the model on extreme cases and then to proceed by showing how GDP growth relates to the two defining components at $t_1$ and $t_2$.

The growth rate is nothing but the first derivative of the function *lnG(t)*. So, we are interested in how the tangent to the curve behaves. Let's first consider the case of the rapid growth in $t_1$. The overall growth rate $g(t_1)$ is the tangent to the curve at point $t_1$. Please notice that if the age specific population is fixed ($dN_s(t_1) = 0$), the inertial growth rate $g_i$ would be the tangent to the *lnG(t)*. Nevertheless, we observe that $dlnN_s(t) > 0$ what results in a rise in the GDP above the inertial level of growth.

The second case is similar, but differs in the direction of the overall growth. Please notice that $g_i(t_1) > g_i(t_2)$ as the attained level of real GDP per capita is higher at $t_2$ and $A/G(t_1) > A/G(t_2)$. Furthermore, the rate of change of age specific population is negative, which leads to the overall negative growth as the GDP declines.

## 2. Annual increment between 1950 and 2011

We start with a revision of the period after 1950. Originally, we estimated the evolution of annual increment before 2003 (Kitov, 2006a). Then data for the period between 2004 and 2007 were added (Kitov, 2009). Here we add four more yearly readings for all involved countries and compare the obtained models with their previous versions. Six years ago, we made a model based assumption that all large deviations from the linear trend in the annual GDP per capita should fade away. This assumption was a successful one and it statistically validates the model of linear GDP growth.

There are many figures since the study is a purely empirical one and the best way to present extensive quantitative results related to time series is to visualize them in form of time history. Nevertheless, our model of real economic growth will also undergo a formal statistical test for normal distribution of residuals. A wider range of specific econometric tools developed for the investigation of real economic growth is presented by Durlauf and co-authors (2005).

Equation (1) suggests that the inertia component of the growth rate of GDP per capita is inversely proportional to the attained level of real GDP per capita, i.e. the observed growth rate should asymptotically approach zero with increasing GDP per capita. On the other hand, the lower is the level the higher is the rate of growth. This inference might be a potential explanation of the empirics behind the concept of economic convergence. The rate of growth must be higher in less developed countries, but the absolute gap in GDP per capita cannot be reduced in the future, unless some non-economic forces will disturb the current status quo.

A cross-country comparison implies that GDP per capita is measured in the same currency units. There are two principal possibilities to reduce national readings of GDP per capita to some common scale: to use currency exchange rates or purchase power parities (PPPs). We use the latter approach and data provided by the Conference Board (2012). For developed countries, two estimates of GDP per capita levels are used: as measured in 2011 US dollars, for which "EKS" purchasing power parities have been applied, and that expressed in 1990 US dollars, with the conversion at Geary Khamis PPPs. These PPPs are obtained from the Organization for Economic Co-operation and Development. Being an improvement on the previous dataset, the "EKS" PPPs are considered as more accurate and reliable.

Only thirteen from thirty four OECD member countries are analyzed. The selection countries all meet the following general criteria: 1) large economy as denominated in US dollars; 2) continuous observations during the period between 1870 and 2011; 3) high level of real GDP per capita. According to the size criterion, quite a few smallest economies were excluded. Furthermore, the third largest economy in the world – Germany is excluded from the data set as it does not meet the requirement for the continuity of time series. The third criterion excluded countries such as Turkey, Poland and other new EU members.

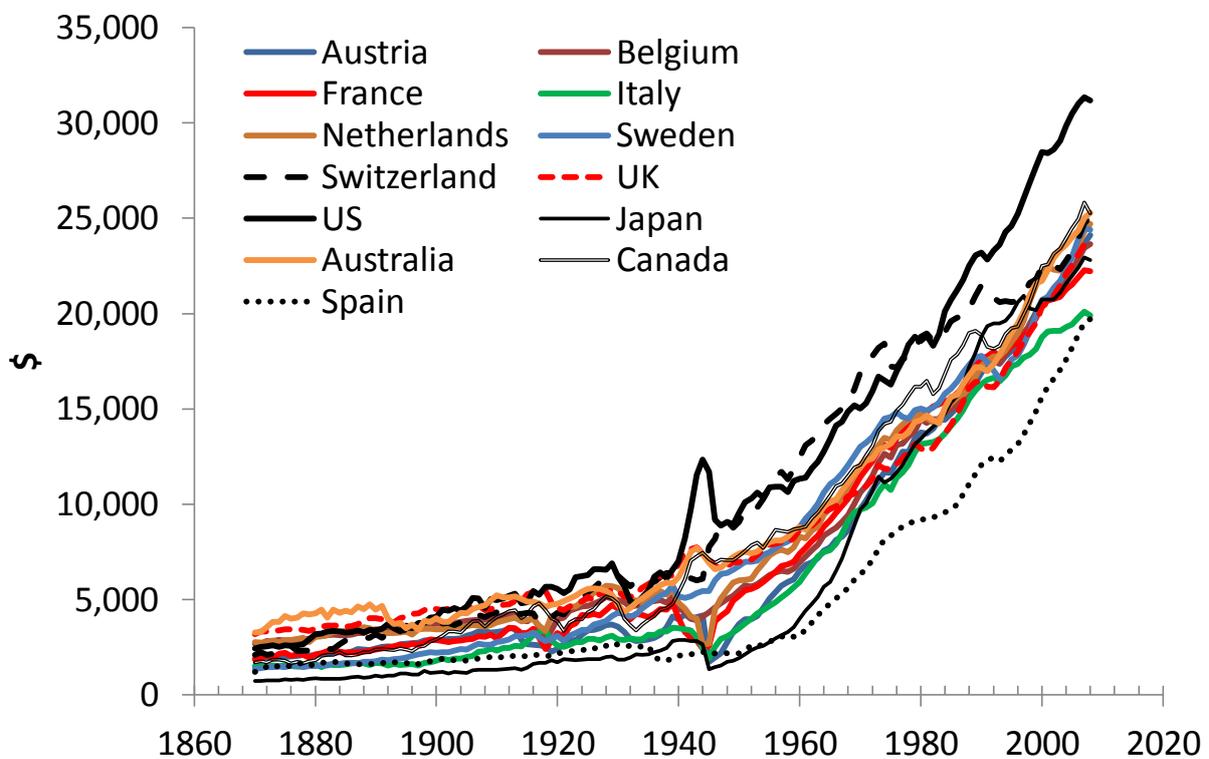

**Figure 2. The evolution of real GDP per capita in thirteen developed countries.**

Figure 2 depicts the overall evolution of real GDP per capita in the countries under study. There is a clear break between 1940 and 1950, with the edges of the break period varying with country. These years include the Second World War and are characterized by an extremely high volatility also associated with somehow abnormal (war-time) functioning of all involved economies. As a result, a significant bias is very likely in the estimates of slopes and their standard errors accompanied by the transition from the reconstructed to measured GDP values. That's why we have skipped these years without any big harm to statistical inferences related to individual segments.

From Figure 2, we calculate regression lines before 1940 and after 1950. Table 1 lists the slopes of regression lines, their standard errors and the ratio of the slopes. The highest slope after 1950 was observed in the US and the lowermost in Switzerland. At the same time, Switzerland has the highest slope before 1940, with Spain evolving at a four time slower speed as defined by US dollars per year. The slope ratio varies from 4.0 in Switzerland to 22.1 in Spain. The slope breaks are statistically significant. We do not test here the hypothesis of exponential growth through the whole period after 1870 since the residuals are not normally distributed before 1940, as shown in Section 3. Also, the null hypothesis of linear trend has a priority since it is simpler and explains two periods better than the hypothesis of exponential growth since 1870.

We start with a revision of the previously obtained estimates. Figure 3 shows the evolution of annual increment in Australia as a function of real GDP per capita. (Notice that (4) implies that one plots dollars vs. dollars with time as a parameter. One can also use (2) and, equivalently, plot the increment vs. time.) There are two regression lines drawn in the Figure: black line corresponds the period between 1950 and 2007, and red line extends the period to 2011. Both regression equations are also shown in the Figure; the lower one is associated with the extended period. Table 2 lists the mean annual increments and their standard deviations.

Our model assumes that in the long run the slope of regression line has to be zero. The residual fluctuations in the annual increment of real GDP per capita depend only on the defining population changes. The number of processes affecting the birth rate, mortality rate and net migration is very large and, according to the Central Limit Theorem, should result in an approximately normal distribution of the deviations. However, these random fluctuations in the defining age population do not presume the unpredictability of real economic growth. On the contrary, the number of nine-year-olds in the USA, which has been proved as the driving force of real GDP, can be counted with any desirable accuracy (Kitov, 2006b). Hence, the growth is predictable.

For Australia, the regression slope is positive for both periods but falls from +0.024 (0.006) to +0.016 (0.006) due to the low rate of growth since 2008 as associated with the global economic crisis. Both slopes are statistically significant with p-value of 0.0003 and 0.005, respectively (see Table 3). Australia is the only example among the biggest economics with a statistically significant positive slope, which is likely related to demographic changes. Nevertheless, our prediction from 2006 that, in the long run, the regression line should be horizontal was valid and the slope has been falling since 2003. We expect the regression line approaching the zero line in the future. As we foresaw six years ago, the healthy growth of the 1990s and the early 2000s has been compensated by a significant fall in GDP. For Australia, Table 2 lists the mean value if $303.2 (1990 US$) with the standard deviation of $257.5.

The observations since 2003 have shown that the positive excursion in the GDP increment was of a temporary character. It is instructive to revisit all estimates for all studied countries in a few years. If the current slump will not end in the next two to five years, the slope may fall below zero.

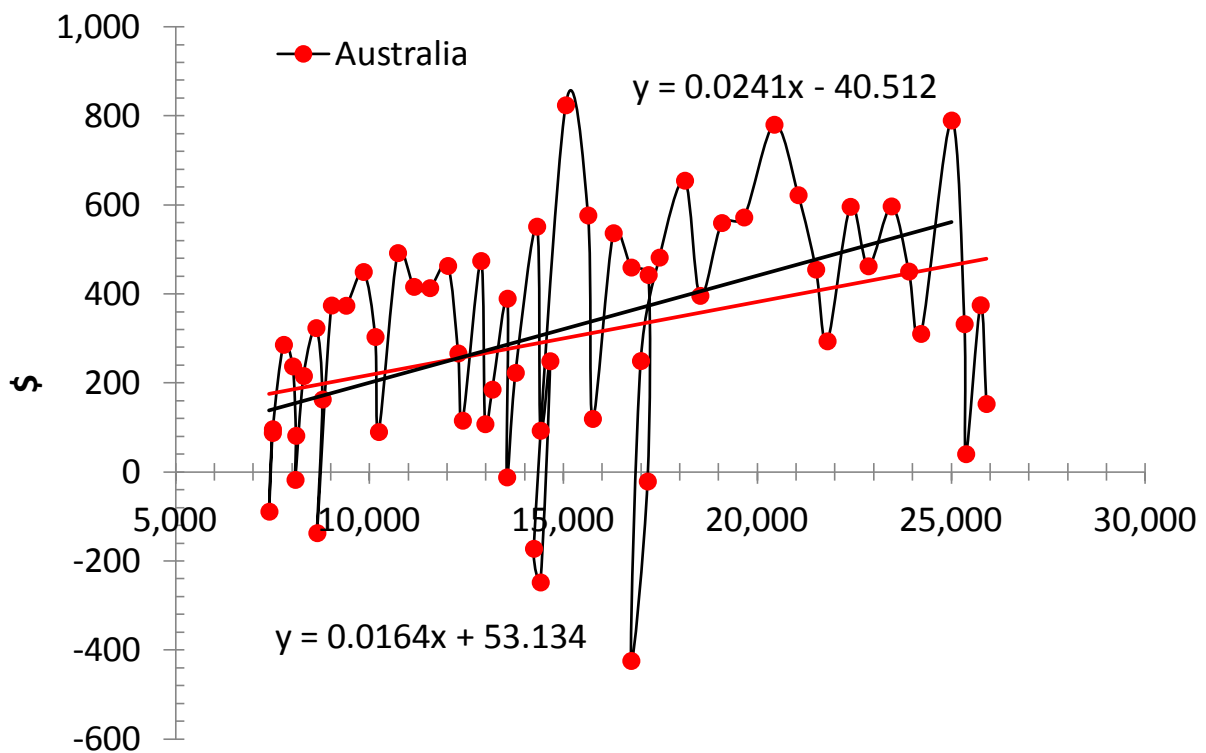

**Figure 3. The annual increment of real GDP per capita (in 1990 US dollars) as a function of real GDP per capita in Australia for the period between 1950 and 2011. The regression (red) line slope is $0.016 per dollar and the mean increment is $303. For the period between 1951 and 2007, the regression (black) line has a larger slope of $0.024 per dollar.**

**Table 1. The slopes of regression lines before 1940 and after 1950, their standard errors and ratio.**

| Country | After 1950 | | Before 1940 | | |
|---|---|---|---|---|---|
| | Slope, $/y | St.Err. | Slope, $/y | St.Err. | Ratio |
| Australia | 310.2 | 6.4 | 25.7 | 2.4 | 12.1 |
| Austria | 349.9 | 3.1 | 21.4 | 1.9 | 16.3 |
| Belgium | 321.3 | 3.5 | 32.3 | 1.5 | 10.0 |
| Canada | 314.0 | 4.9 | 48.8 | 2.6 | 6.4 |
| France | 298.1 | 3.9 | 38.2 | 1.7 | 7.8 |
| Italy | 286.2 | 5.6 | 30.5 | 1.4 | 9.4 |
| Japan | 348.1 | 10.2 | 24.6 | 0.9 | 14.1 |
| Netherlands | 319.5 | 4.8 | 38.9 | 2.0 | 8.2 |
| Spain | 319.5 | 4.8 | 14.5 | 1.1 | 22.1 |
| Sweden | 299.0 | 6.5 | 52.9 | 2.1 | 5.7 |
| Switzerland | 247.2 | 6.1 | 61.4 | 2.0 | 4.0 |
| UK | 282.6 | 6.5 | 39.4 | 1.6 | 7.2 |
| US | 387.7 | 6.7 | 60.9 | 2.4 | 6.4 |

**Table 2. Mean annual increments and their standard deviations**

| Country | Mean 1870 | St.dev. 1870 | Mean 1950 | St.dev. 1950 | mean1950/ mean1870 |
|---|---|---|---|---|---|
| Australia | 41.3 | 221.7 | 303.2 | 257.5 | 7.3 |
| Austria | 30.0 | 156.1 | 344.2 | 272.9 | 11.5 |
| Belgium | 26.7 | 188.5 | 303.9 | 264.9 | 11.4 |
| Canada | 52.5 | 239.6 | 295.2 | 353.2 | 5.6 |
| France | 31.0 | 216.5 | 272.2 | 223.7 | 8.8 |
| Italy | 28.7 | 123.5 | 242.5 | 277.6 | 8.6 |
| Japan | 30.5 | 81.7 | 297.3 | 412.7 | 9.7 |
| Netherlands | 29.6 | 192.3 | 307.2 | 305.4 | 10.4 |
| Spain | 12.5 | 108.8 | 240.7 | 236.0 | 19.3 |
| Sweden | 54.6 | 138.8 | 317.5 | 395.1 | 5.8 |
| Switzerland | 61.4 | 167.1 | 271.7 | 378.6 | 4.4 |
| United Kingdom | 52.4 | 164.4 | 253.1 | 315.8 | 4.8 |
| United States | 65.2 | 283.0 | 350.3 | 431.3 | 5.4 |

The next country is Austria as shown in Figure 4. The average increment since 1950 is $344.2 with the standard deviation of $272.9. The mean annual increment is well above that for Australia (more than 10% per year). A prominent feature is an almost horizontal regression line with a slope of +0.006 (0.006); since the relevant p-value is 0.26 one can not reject the null hypothesis of a zero slope. For the period before 2007, the slope is slightly larger +0.012 (0.005) and p-value is 0.014. Therefore, it is possible to conclude that the relevant specific age population in Austria in (1) has changed only marginally during from 1950 to 2011, i.e. it did not affect the inertial growth in the long run. It is illustrative that the increment fell down in absolute terms by $944 in 2009 what is equivalent to $1288 relative to the mean value since 1950. This is a dramatic fall. The largest absolute rise of $842 was measured in 2007.

Table 3. Regression of annual increments: slopes, their standard errors, t-statistics, and p-values for both studied periods

| Country | 1870-1940 | | | | 1950-2011 | | | |
|---|---|---|---|---|---|---|---|---|
| | Slope | St.Err. | t Stat | p-value | Slope | St.Err. | t Stat | p-value |
| Australia | 0.071 | 0.040 | 1.773 | 0.081 | 0.016 | 0.006 | 2.953 | 0.005* |
| Austria | 0.057 | 0.034 | 1.655 | 0.102 | 0.006 | 0.006 | 1.137 | 0.260 |
| Belgium | 0.025 | 0.032 | 0.773 | 0.442 | 0.007 | 0.006 | 1.240 | 0.220 |
| Canada | 0.037 | 0.026 | 1.407 | 0.164 | 0.006 | 0.008 | 0.777 | 0.440 |
| France | 0.030 | 0.031 | 0.953 | 0.344 | -0.006 | 0.005 | 1.077 | 0.286 |
| Italy | 0.035 | 0.022 | 1.607 | 0.113 | -0.013 | 0.007 | 1.947 | 0.056 |
| Japan | 0.060 | 0.017 | 3.456 | 0.001 | -0.010 | 0.008 | 1.164 | 0.249 |
| Netherlands | 0.021 | 0.027 | 0.783 | 0.436 | 0.006 | 0.007 | 0.888 | 0.378 |
| Spain | 0.020 | 0.038 | 0.527 | 0.600 | 0.002 | 0.006 | 0.298 | 0.767 |
| Sweden | 0.032 | 0.014 | 2.221 | 0.030 | 0.017 | 0.009 | 1.856 | 0.068 |
| Switzerland | 0.017 | 0.015 | 1.103 | 0.274 | -0.005 | 0.011 | 0.481 | 0.632 |
| UK | 0.057 | 0.022 | 2.522 | 0.014 | 0.007 | 0.008 | 0.902 | 0.370 |
| US | 0.032 | 0.026 | 1.236 | 0.221 | 0.006 | 0.008 | 0.776 | 0.441 |

*the null is rejected at 1% level

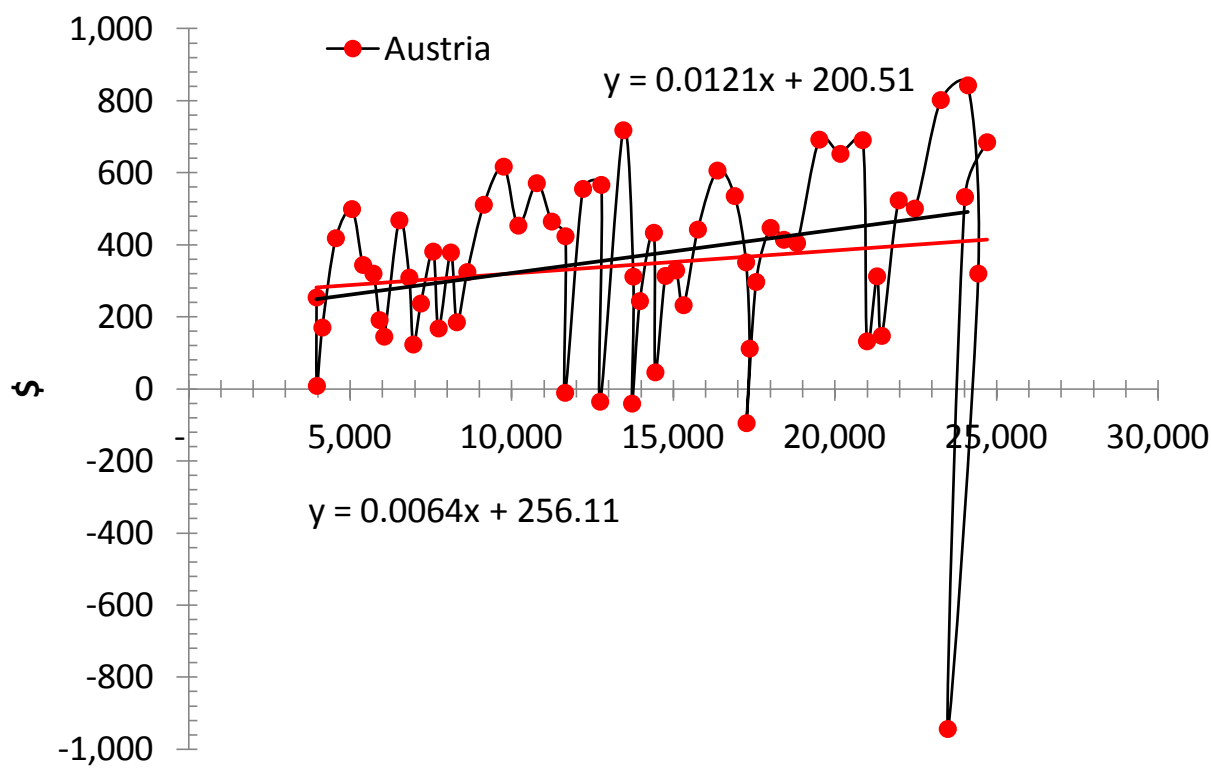

**Figure 4. The annual increment of real GDP per capita (in 1990 US dollars) as a function of real GDP per capita in Austria for the period between 1950 and 2011.**

The average increment in Belgium is $304 with the standard deviation of $265. Both regression lines (Figure 5) are characterized by positive slopes close to those in Austria; the early regression line being also steeper. The current slope is not statistically distinguishable from zero. The largest deviation from the mean value was measured in 2009: -$1000. As in the other countries, the negative deviations are usually sharp and somehow compensating relatively long periods of weaker positive growth. The twenty years before 2008 were good for Belgium. As for Austria, we expect a further slope decline in the near future.

The case of Canada in Figure 6 is rather similar to that of Austria and Belgium despite it was closer to Australia in 2003. The mean increment is $295 (±$353). The current slop of +0.006 (0.008) is characterized by p-value 0.44 and thus only insignificantly differs from zero. A striking feature of the Canadian fluctuations is three negative peak amplitudes: -$988 in 1982, -$905 in 1991, and -$1196 in 2009.

For France (Figure 7), we first notice a negative slope: -0.006 (0.005). Despite its statistical insignificance (the corresponding p-value is 0.28) the weak negative linear trend in France is very illustrative. There is no exponential growth in real GDP per capita as Solow model suggests. The mean increment is $272 that is slightly lower than in Austria and Belgium. However, the fluctuations in France are characterized by a smaller scattering with the standard deviation of only $224. We expect the negative trend to survive through the next few years together with the overall turmoil in major European economies.

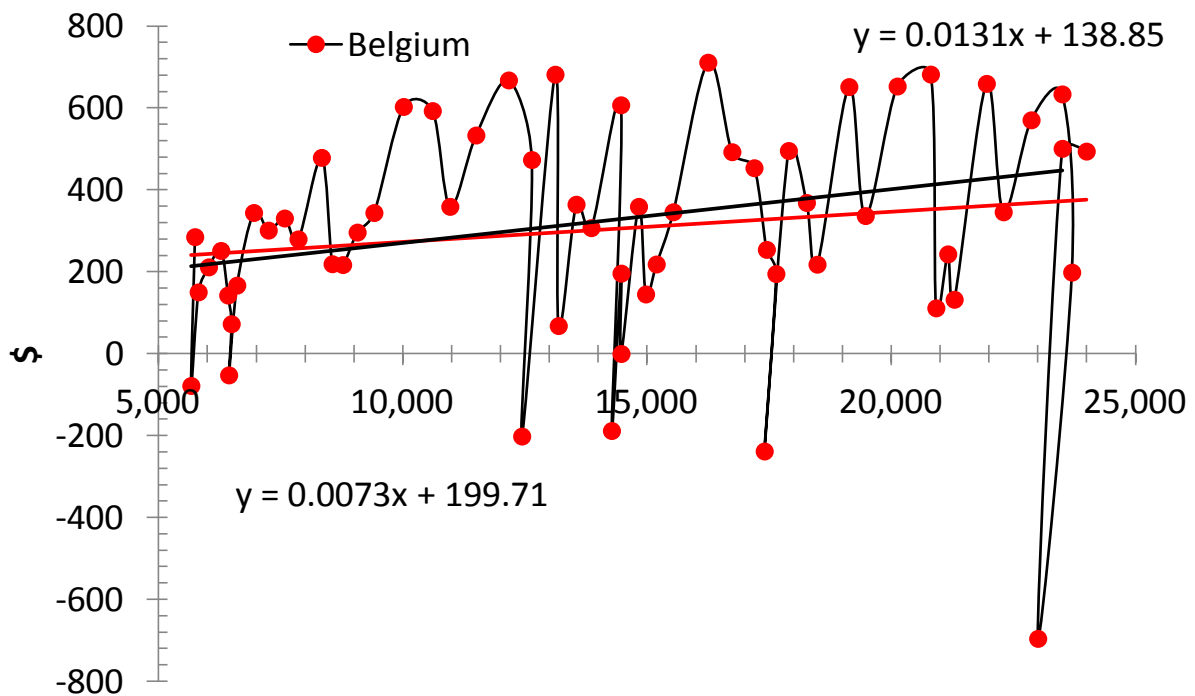

**Figure 5. The annual increment of real GDP per capita (in 1990 US dollars) as a function of real GDP per capita in Austria for the period between 1950 and 2011.**

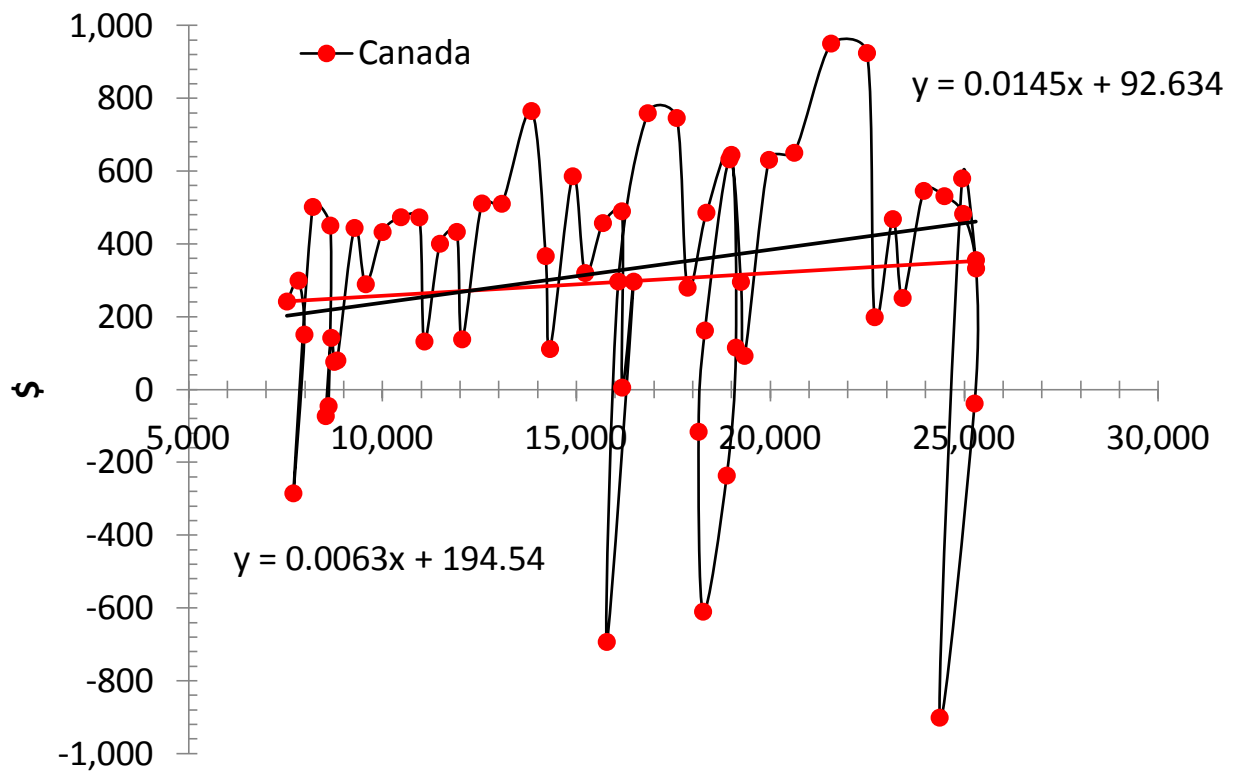

**Figure 6. The annual increment of real GDP per capita (in 1990 US dollars) as a function of real GDP per capita in Canada for the period between 1950 and 2011.**

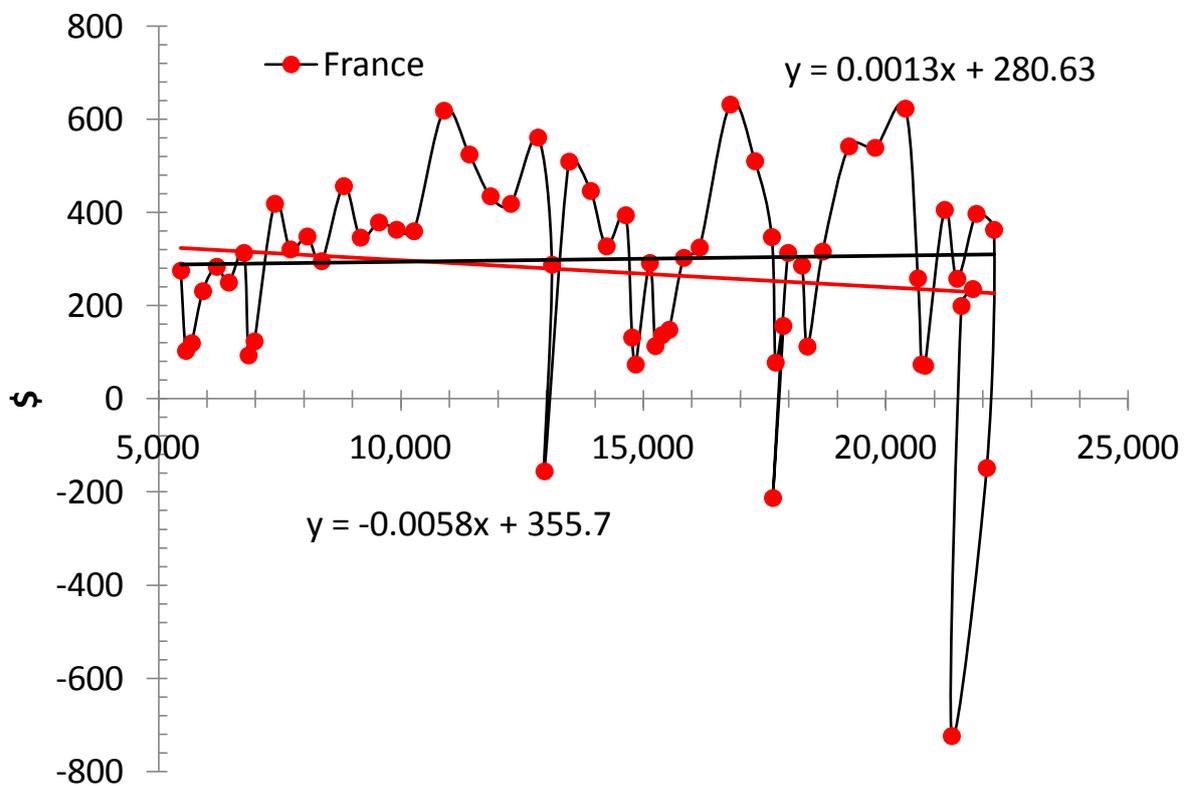

**Figure 7. Same as in Figure 3 for France. The mean increment is $272 (σ=$224).**

The next two countries are Italy and Japan as displayed in Figures 8 and 9, respectively. They are not only close alphabetically but both demonstrate negative trends in GDP since 1950: -0.013 and -0.0097, respectively. The mean increment for Italy is very low $242 with the standard deviation of $278. Japan is characterized by a larger mean of $298 and a much higher volatility of $413. An outstanding feature of the Japanese curve is the oscillation around $20,000 since 1990. This is a compensation of the previous growth related to the rise in the number of 18-year-olds (Kitov and Kitov, 2010). Since 2010, the negative influence of the population pyramid will be fading away and the Japan economy will be growing at an inertial rate with $A$=$297. In 2011, the inertial growth rate of real GDP per capita is $A/G$=$297/$20054=0.0015 y$^{-1}$.

The Kingdom of the Netherlands has a slight positive trend with the slope of $0.006 (0.007) per dollar. The null hypothesis of a zero slope cannot be rejected for p-value of 0.38. The past four years were not good for the economy and the trend has fallen since our previous estimate made in 2009. With the mean increment of $307 and standard deviation of $305, the Netherlands is in the middle of both distributions. One should not exclude any further decline in the slope.

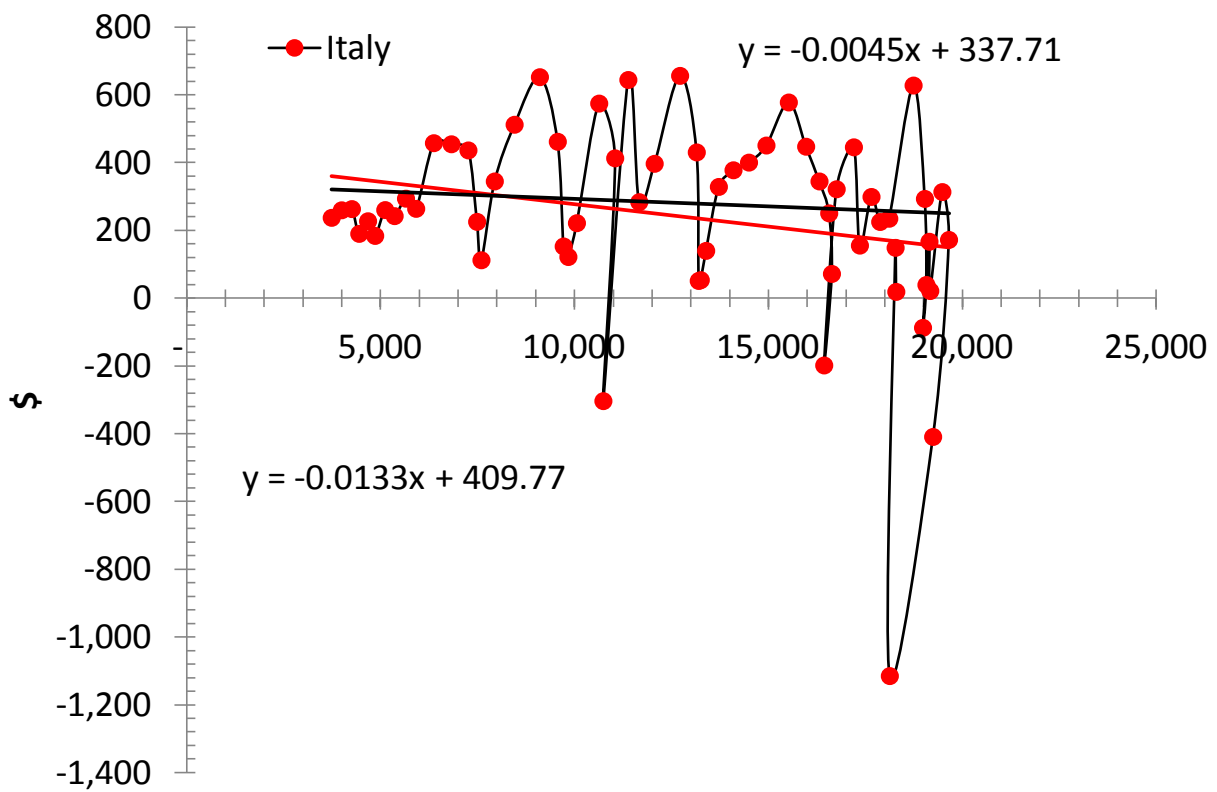

**Figure 8. Same as in Figure 3 for Italy. The mean increment is $242 (σ=$278).**

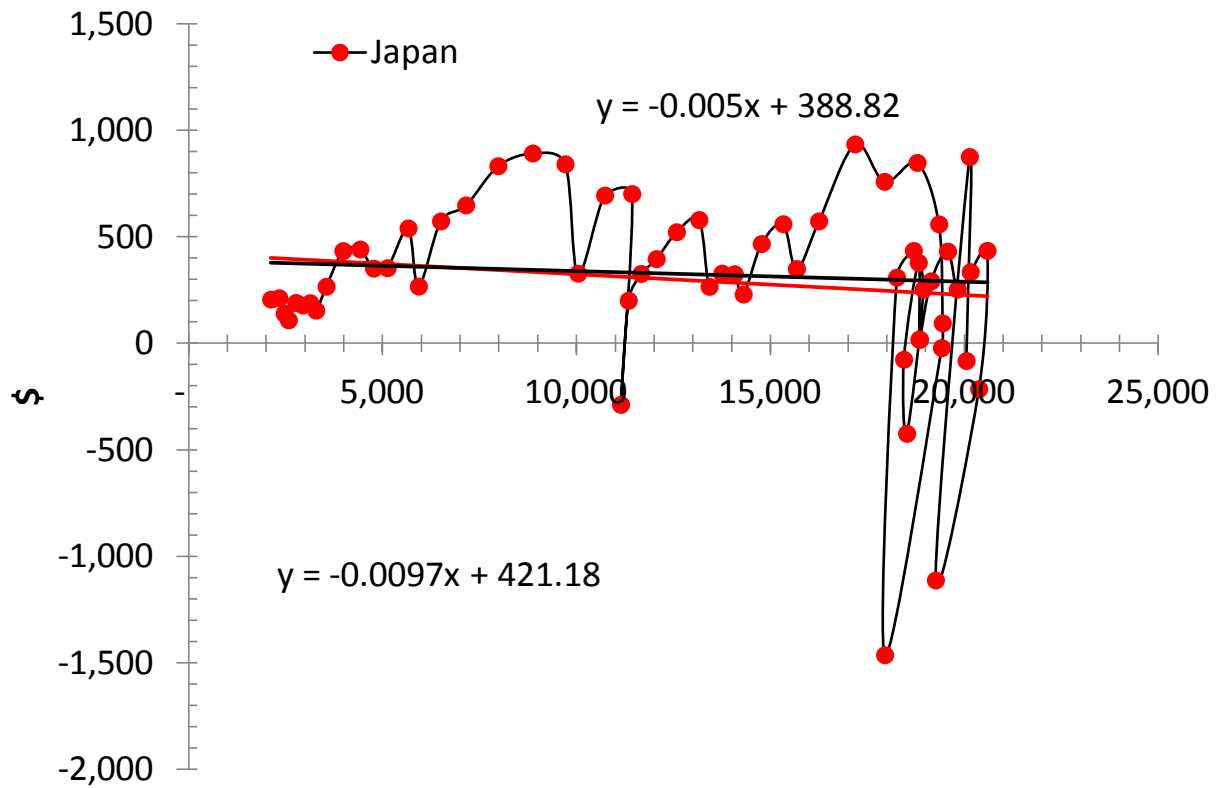

**Figure 9. Same as in Figure 3 for Japan. The mean increment is $297 (σ=$413).**

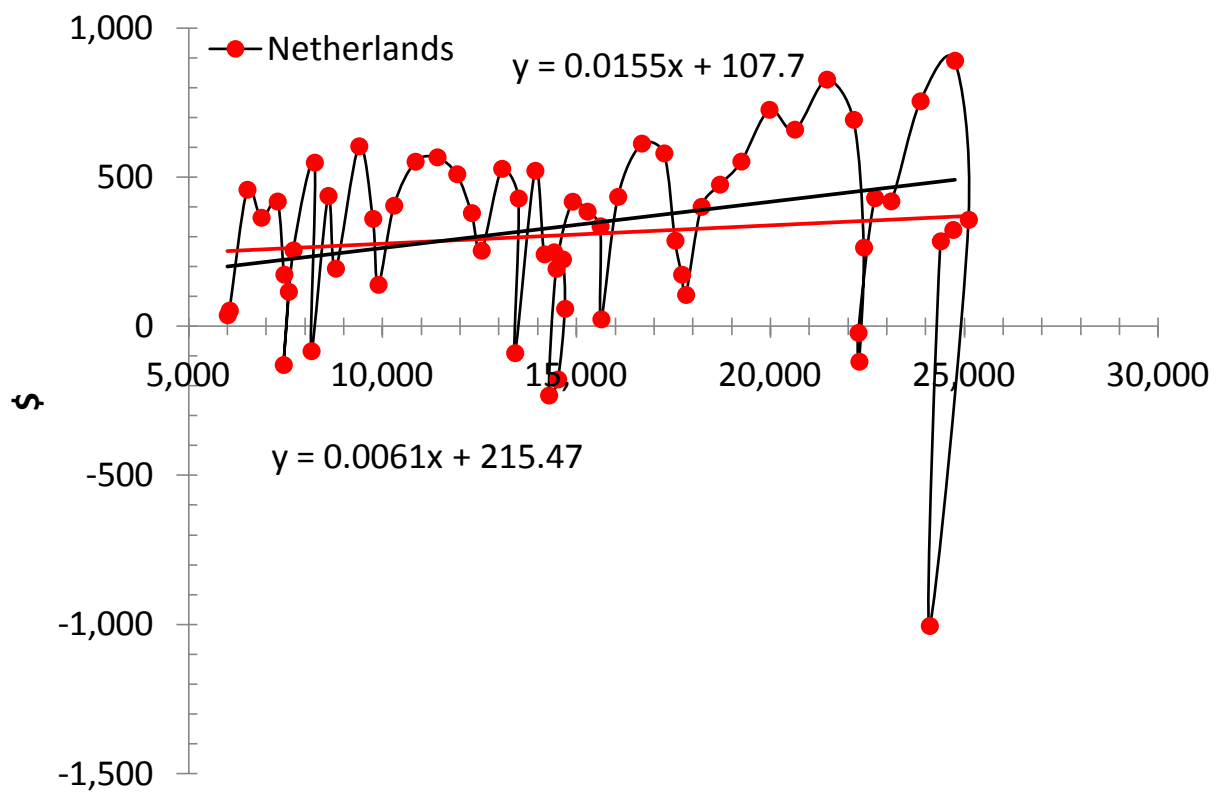

**Figure 10. Same as in Figure 3 for the Netherlands. The mean increment is $307 (σ= $305).**

Spain in Figure 11 has the smallest mean increment of $241 with the standard deviation of $236 which is also small among the studied countries. There were two periods of stagnation in Spain: near $3000 and $10000. This is likely the reason of the weak overall growth since 1950. Currently, the slope of +0.001 (0.006) is indistinguishable from zero. In 2012, a new recession was announced. The Spanish long term trend will likely sink below the zero line and stay there for some time.

Sweden had been demonstrating a healthy and sustainable growth since 1950 before the 2008/2009 recession. Then, a record fall of $1340 was measured. With the mean annual increment of $318, accompanied by a standard deviation of $395, the regression line in Figure 12 has a slope estimate of +0.017 (0.009), which is not significant at the 5% level for p-value 0.064.

Switzerland (Figure 13) has a decreasing increment with a very small slope of -0.005 (0.011) which is statistically insignificant. The same slope was estimated in 2009. The average increment is rather small $271 with a larger standard error of $379. There is no sign of the Swiss GDP to rise faster than in the past and thus there is no expectation of a positive slope in the increment curve. Historically, Switzerland has quite a few years of negative growth together with many more successful episodes of growth.

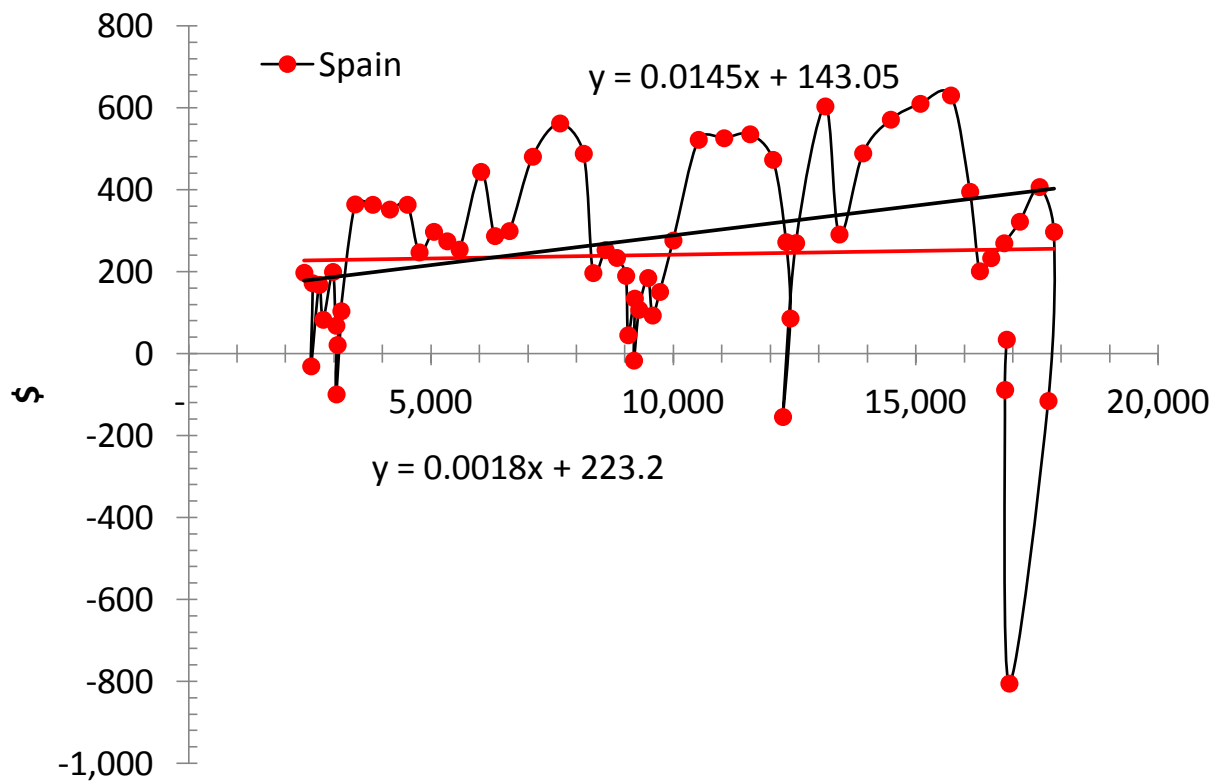

**Figure 11. Same as in Figure 3 for Spain. The mean value is $241 (σ=$236).**

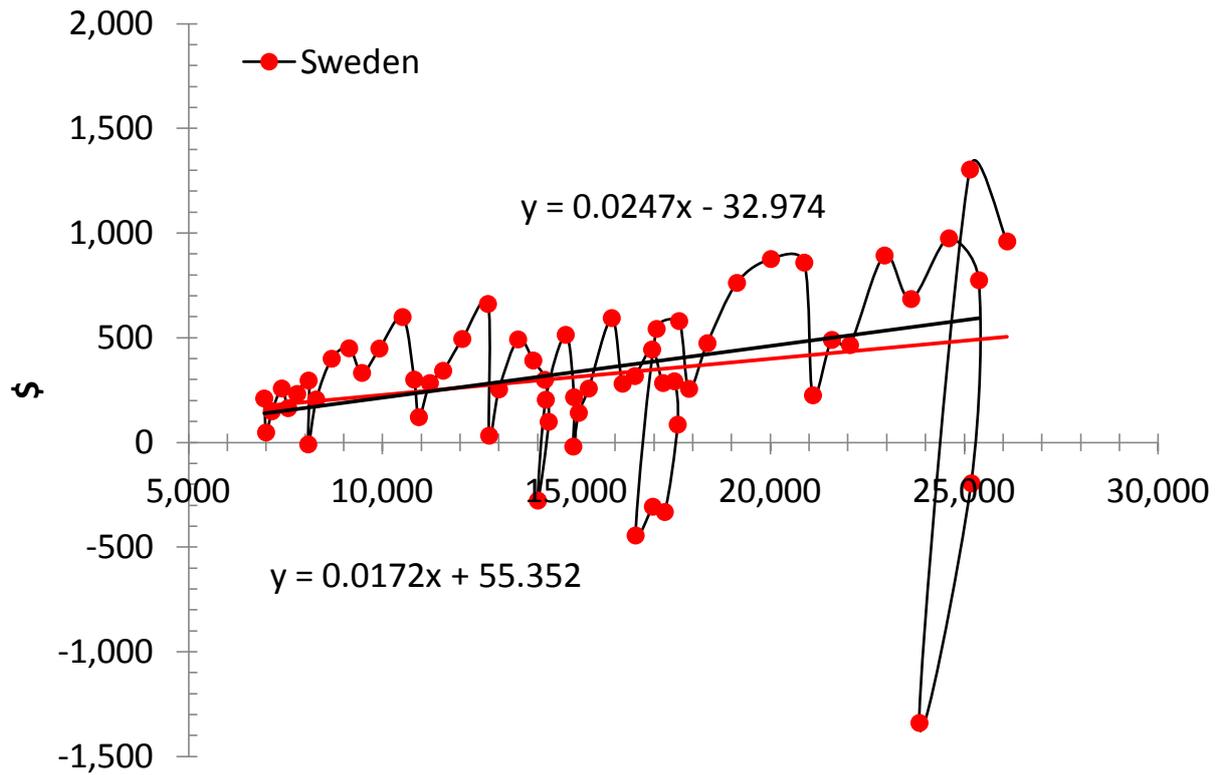

**Figure 12. Same as in Figure 3 for Sweden. The mean value is $378 (σ=$395).**

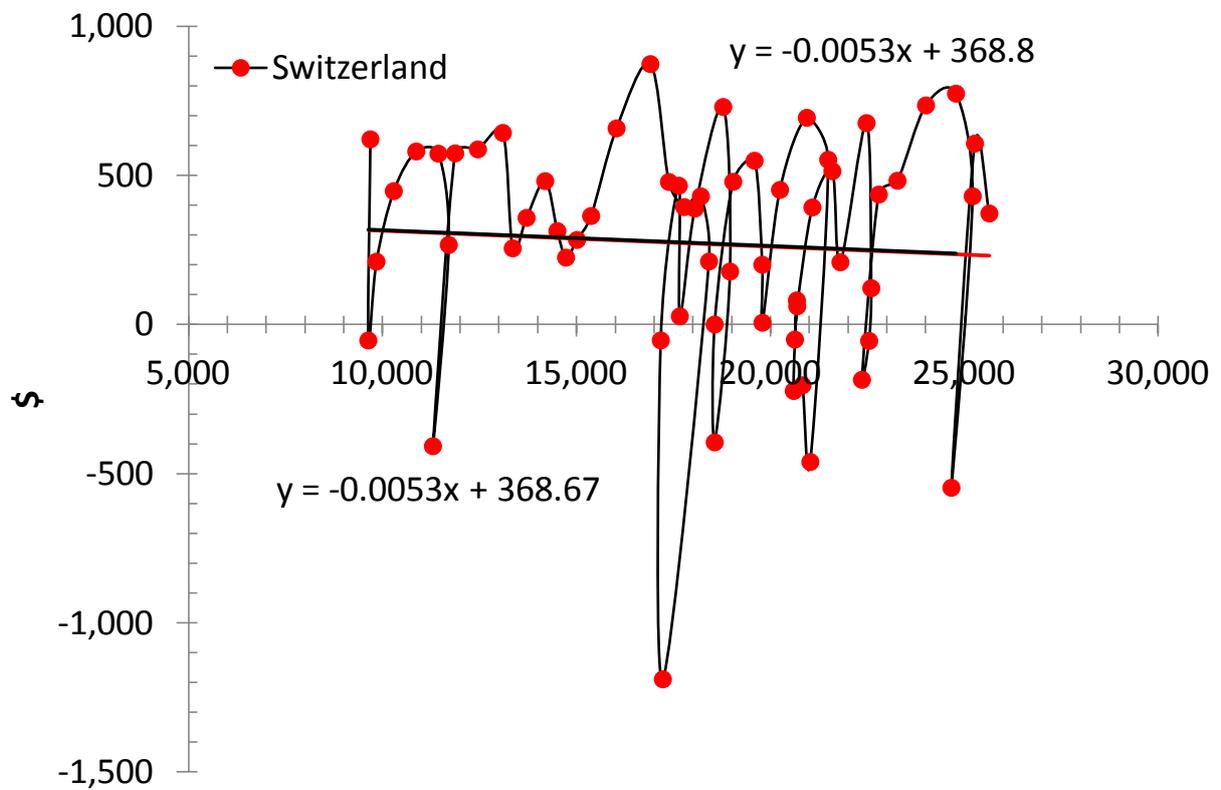

**Figure 13. Same as in Figure 1 for Switzerland. The mean value is $378 (σ=$379).**

The United Kingdom and the United States differ only in the mean increments and standard deviations: $253 (σ=$316) and $350 (σ=$431), respectively. Both regression lines from 1950 to 2011 are characterized by a negligible positive slope. Four years ago, these lines had steeper and statistically significant slopes, but the history of the current crisis has added a few negative values to these series. This might be considered as a fundamental property of inertial economic growth: in the long run no large positive deviations are possible. However, one can not exclude a lengthy underperformance which may be caused by non-economic forces. Spain is likely an example of such underperformance.

The US trend was well explained four years ago by the change in the nine-year-old population (Kitov, 2009). When corrected for the overall nine-year-olds change between 1950 and 2007, the US mean value was only $462 (2008 US dollars). The growth in the number of 9-year-olds has reverted since the early 2000s and has now a negative input in GDP per capita. Unfortunately, the UK statistical agencies do not provide accurate population estimates for the entire period since 1950, but from the mean value one can assume that there was no significant increase in the specific age population.

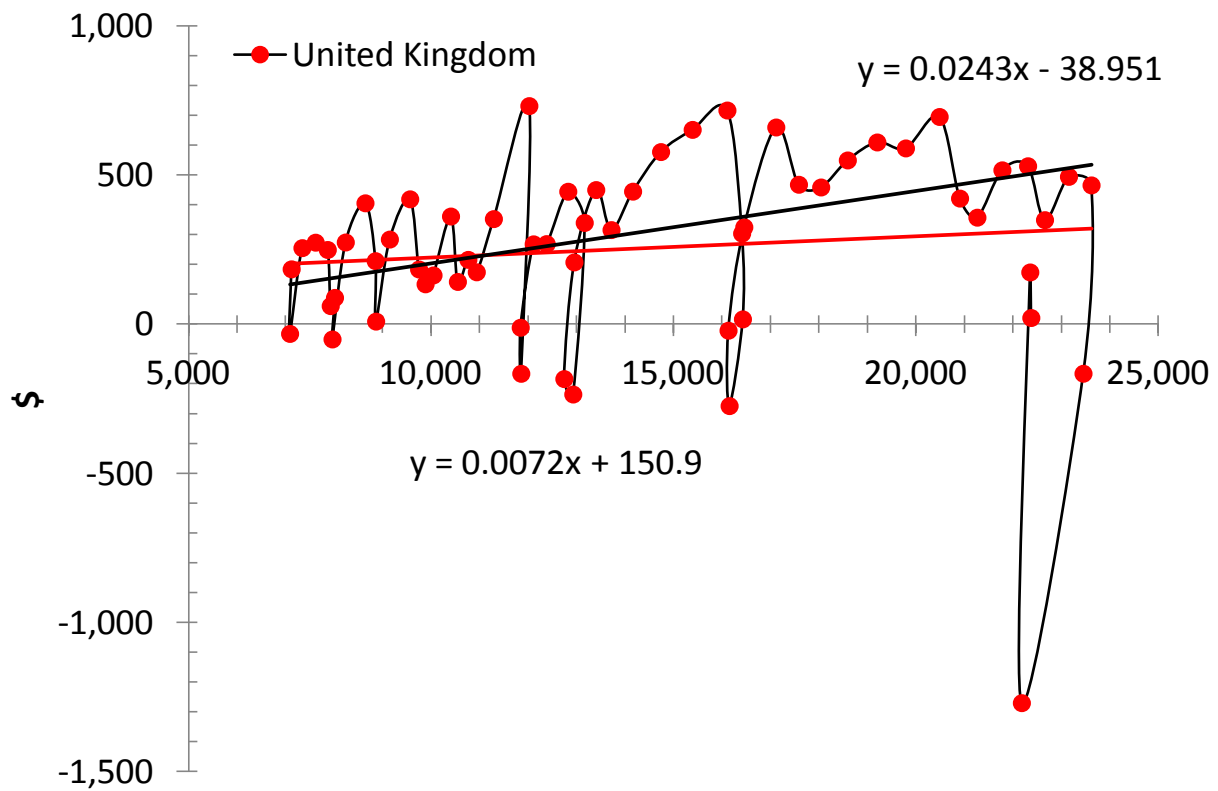

**Figure 14. Same as in Figure 1 for the United Kingdom. The mean value is $316.**

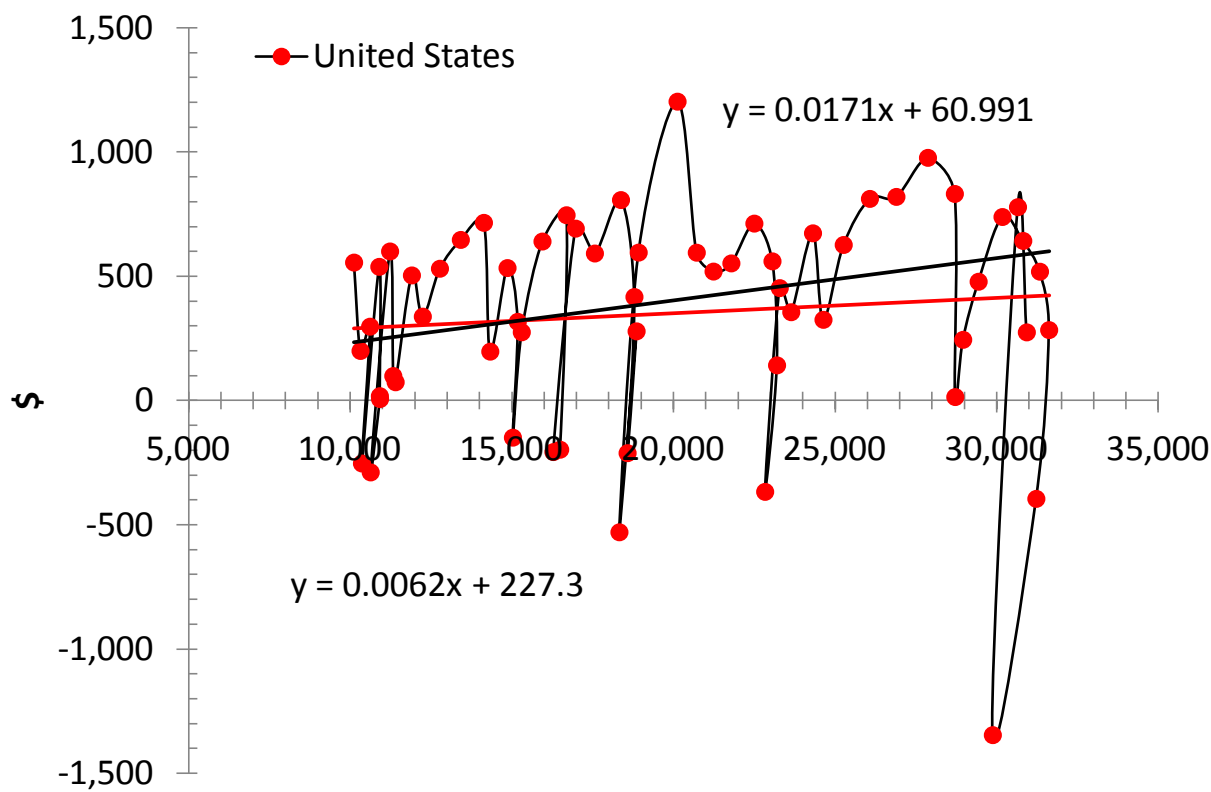

**Figure 15. Same as in Figure 1 for the United States. The mean value is $350.**

As we expected four years ago, all thirteen countries have shown a decline in regression slopes associated with the current economic slump. The null hypothesis of a zero slope can not be rejected for almost all countries except Australia where the slope has been also falling since we first estimated it in 2006. In addition, there are several countries with a slightly negative trend: France, Italy, Japan, and Switzerland.

Despite the very close slope estimates, real GDP per capita in the studied countries has different mean increment and standard deviations. The largest mean increment belongs to the United States ($350) and the smallest is measured in Spain ($241). The highest scattering is also observed in the US ($431) with the lowermost dispersion in France ($223). The mean annual increment has crucial significance for the long term growth: even a negligible difference of $10 in 1950 cumulates to $600 in 2011. Therefore, the difference of $110 (Spain and the US) is a catastrophic one. Economically, these countries diverge in time. In part, these differences might be related to definitional problems in the PPP conversion methodology. But we can not exclude that a bigger part of the difference in the mean increment is a genuine one.

As discussed above, the residual increments should be normally distributed if they are related to stochastic processes in the specific age population. Therefore, these differences between the measured and mean annual increments were tested for normal distribution. We have calculated a frequency distribution of the residual increments in $200-wide bins for the original and the demeaned data sets. Figure 16 depicts both distributions. When applied to the whole data set of 793 demeaned observations from all thirteen countries, the Shapiro-Francia test for normality gives p-value < 0.00001, i.e. the null hypothesis of normality can be rejected at 1% significance level. This is a common situation in economic time series with heavy tails, as Figure 16 shows. There are a few outliers in the series beyond ±$800. They introduce a significant bias and the null of normality is rejected. One can remove the outliers (19 in total) and test the residual set for normality. The Shapiro-Francia test gives p-value > 0.032, i.e. the null of normality cannot be rejected at the same significance level. Hence, the residuals are likely taken from a normally distributed population.

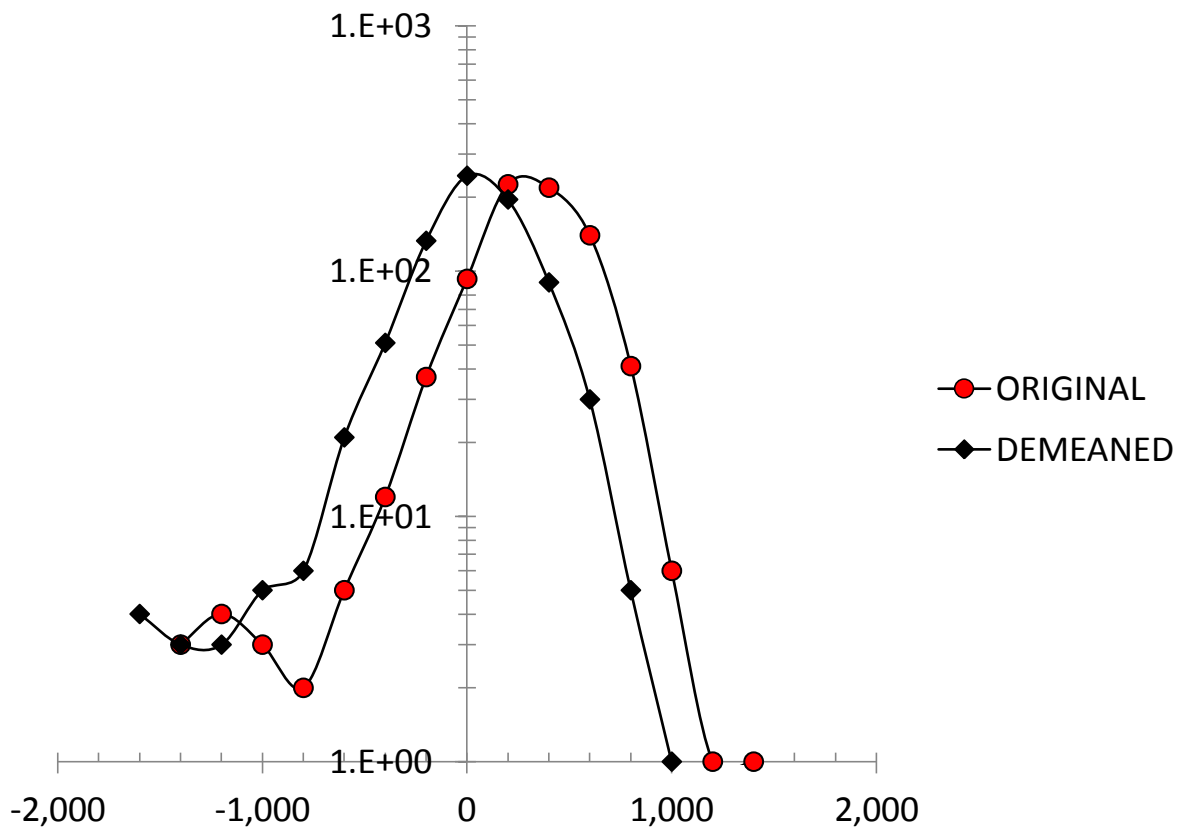

**Figure 16. Frequency distribution of annual increments from 1950 to 2011: original and demeaned.**

In this Section, the results from (Kitov, 2009) have been revisited. All new data support the assumption of inertial economic growth and thus the constancy of annual increment in real GDP per capita in developed economies. The validation process based on new data is the most reliable one. In macroeconomics, it takes a while, however. The aforementioned historical data set is a less reliable one but gives more data when it's possible to obtain in one's lifetime. The next Section analyses these historical estimates using the same approach and statistical tools.

## 3. Annual increment between 1870 and 1940

All annual GDP estimates in this Section are borrowed from the data set complied by Angus Maddison. We retain the same countries as in Section 2 but display the evolution of GDP increment in time between 1871 and 1940, as defined in (2), instead of its closed form (4). The period ends with 1940 because of clear turbulence in the global economy during the WWII and the next several years. Figure 2 also suggests a break in all time series with the mean increment changing by a factor of 4.4 for Switzerland to 19.3 for Spain (see Table 2).

Figure 17 shows all regression lines together with the corresponding equations. According to (4), the slope is measured in (1990) US dollars per year. Table 4 lists corresponding slopes, their standard errors, t-statistic, and p-values. All slopes are insignificant and the null of a zero slope cannot be rejected. Tables 3 and 4 prove a fundamental finding: the growth of real GDP per capita in the biggest developed countries is a linear function of time. Hence, real economic growth is not exponential as presumed in Solow growth model. The nature of the break between 1940 and 1950 is likely related to definition of real GDP.

**Table 4. Time regression of the annual increments between 1870 and 1940.**

|  | *Coefficients* | *Standard Error* | *t Stat* | *p-value* |
|---|---|---|---|---|
| Australia | 0.356 | 1.320 | 0.270 | 0.788 |
| Austria | 0.354 | 0.929 | 0.380 | 0.705 |
| Belgium | -0.478 | 1.122 | -0.427 | 0.671 |
| Canada | 0.708 | 1.425 | 0.497 | 0.621 |
| France | -0.216 | 1.290 | -0.168 | 0.867 |
| Italy | 0.692 | 0.731 | 0.946 | 0.347 |
| Japan | 1.061 | 0.470 | 2.259 | 0.027 |
| Netherlands | -0.217 | 1.145 | -0.190 | 0.850 |
| Spain | -0.680 | 0.643 | -1.058 | 0.294 |
| Sweden | 1.403 | 0.810 | 1.733 | 0.088 |
| Switzerland | 0.585 | 0.993 | 0.588 | 0.558 |
| UK | 1.257 | 0.967 | 1.299 | 0.198 |
| US | 0.552 | 1.685 | 0.328 | 0.744 |

Figure 18 displays the frequency distribution of residuals, both original and demeaned. Unlike the residuals of the measured GDP in Section 2, the artificially reconstructed and subjectively estimated past GDP values reveal an example of exponential distribution rather than a normal one. This confirms the artificiality of the estimates made for the "pre-GDP era" when the concept of Gross Domestic Product did not exist at all.

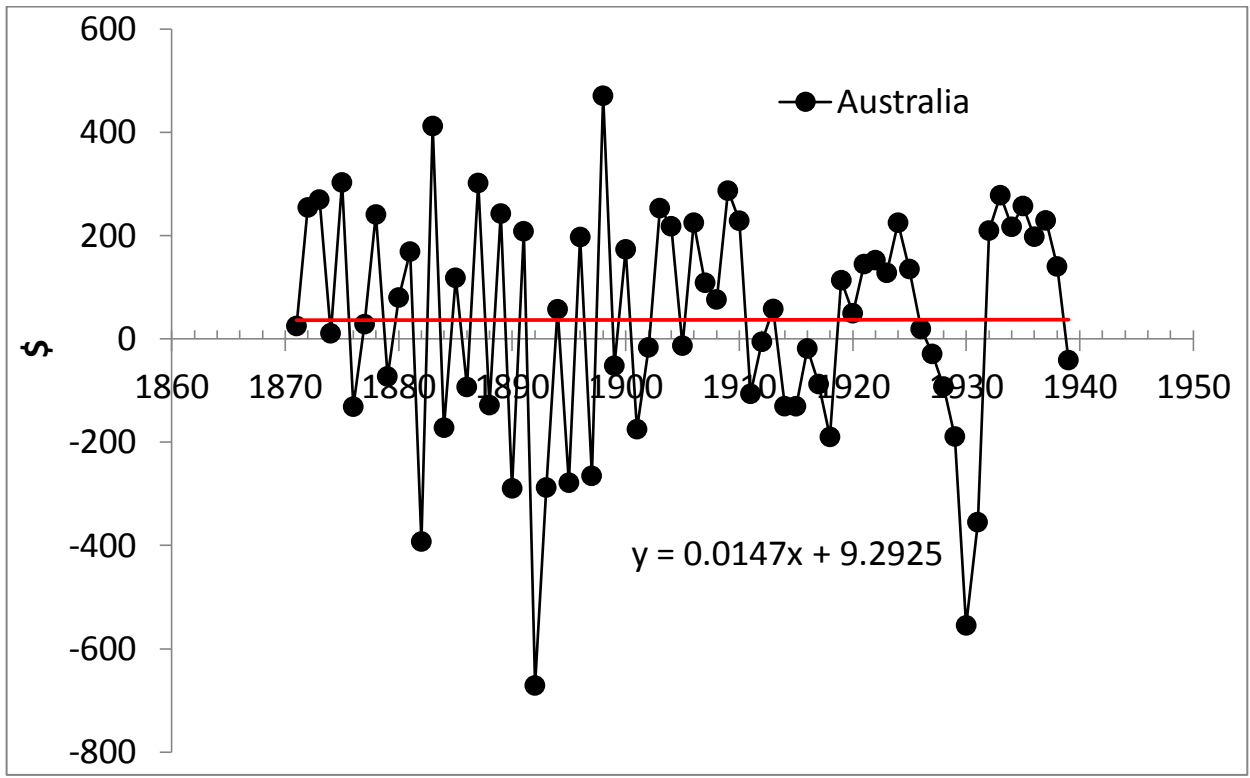

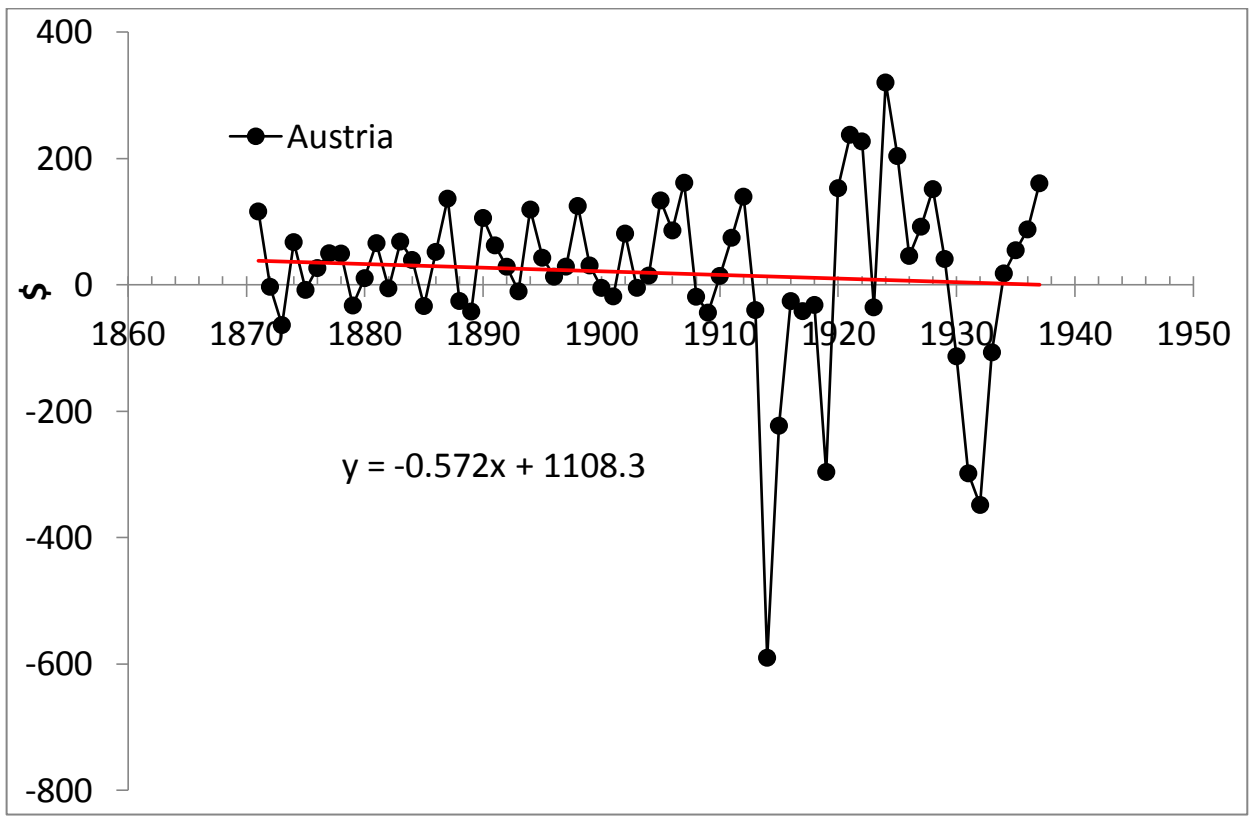

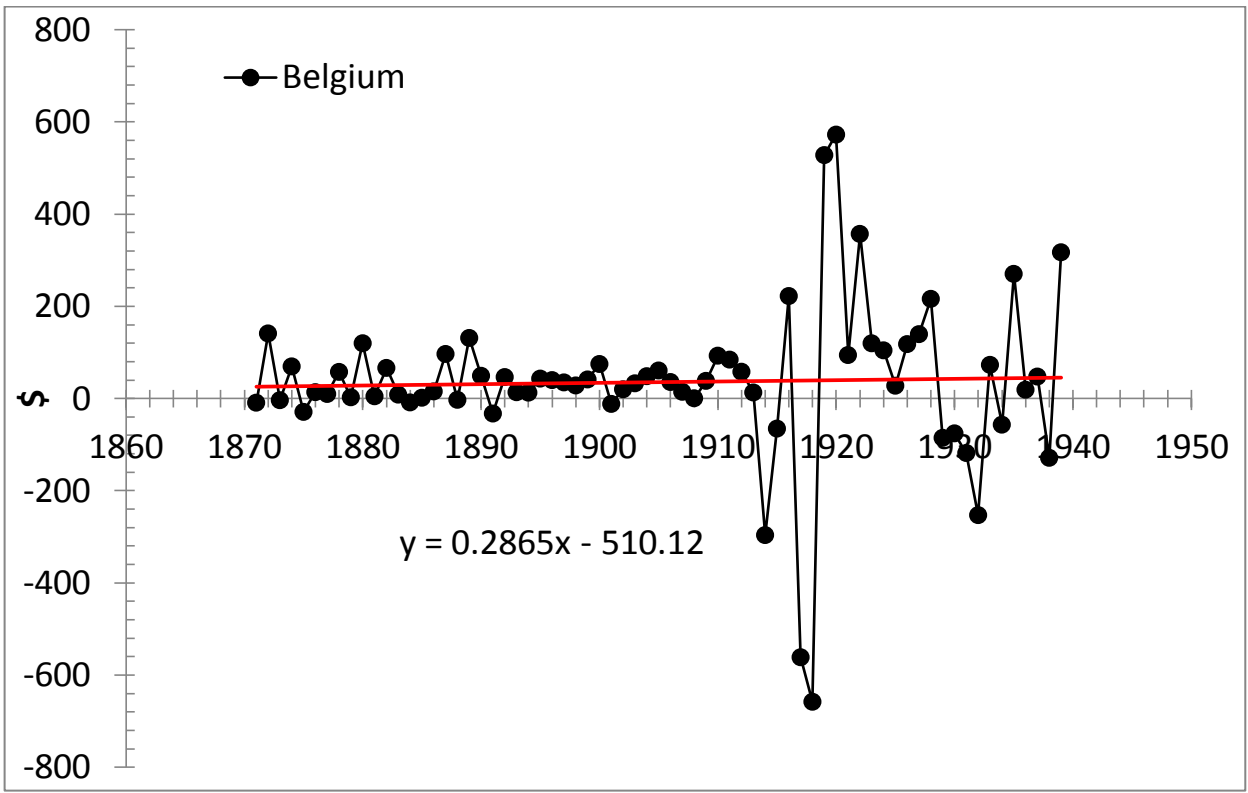
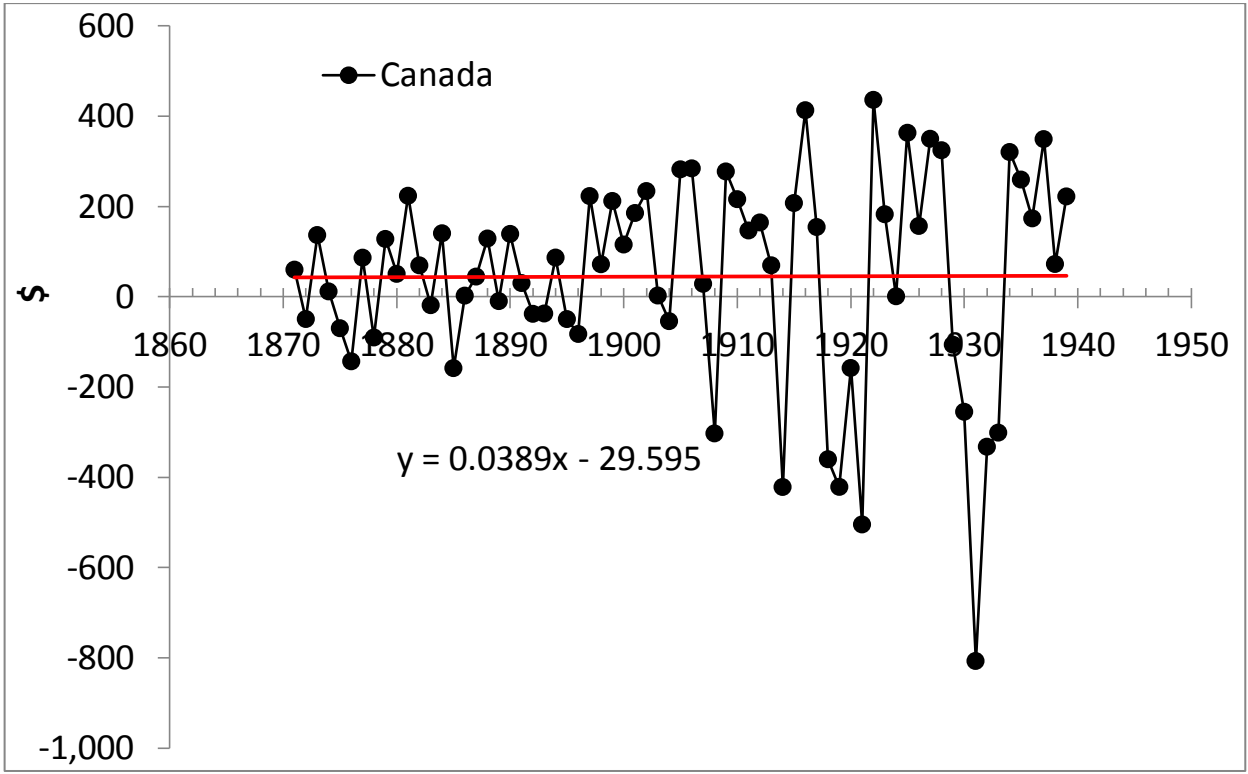

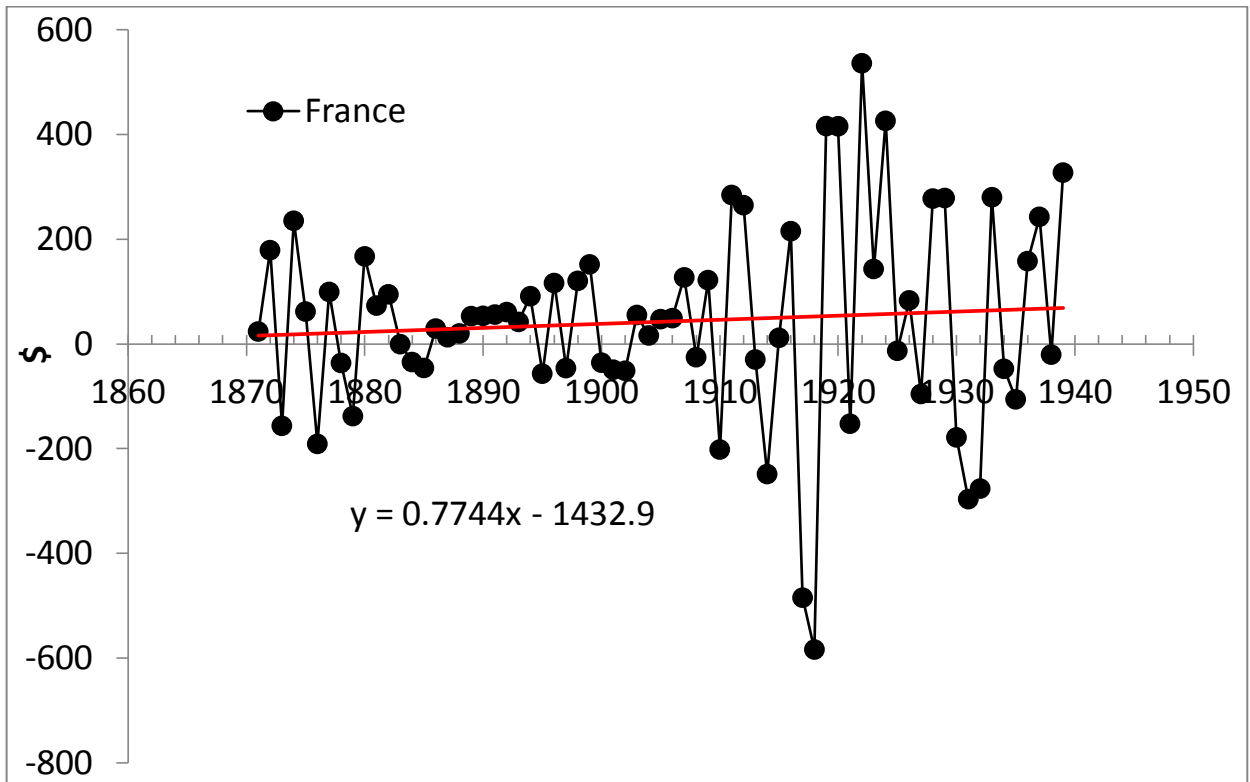

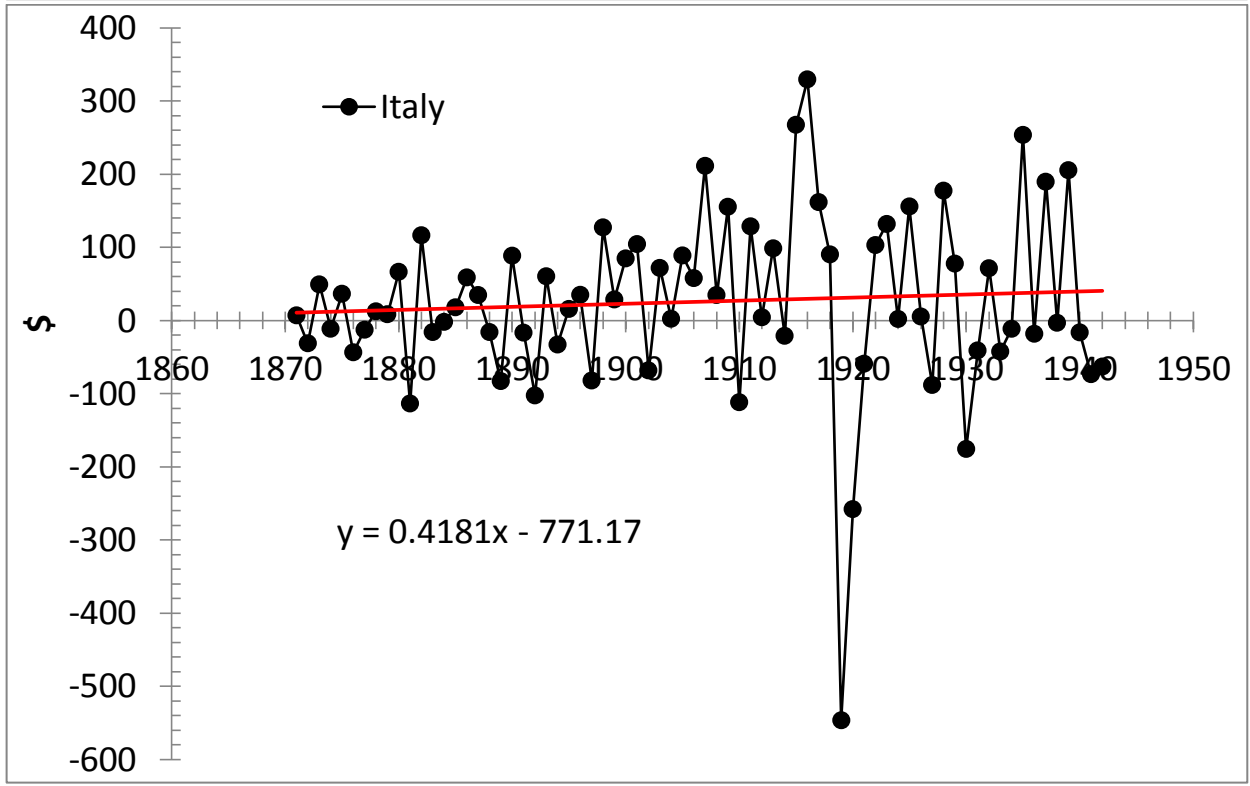

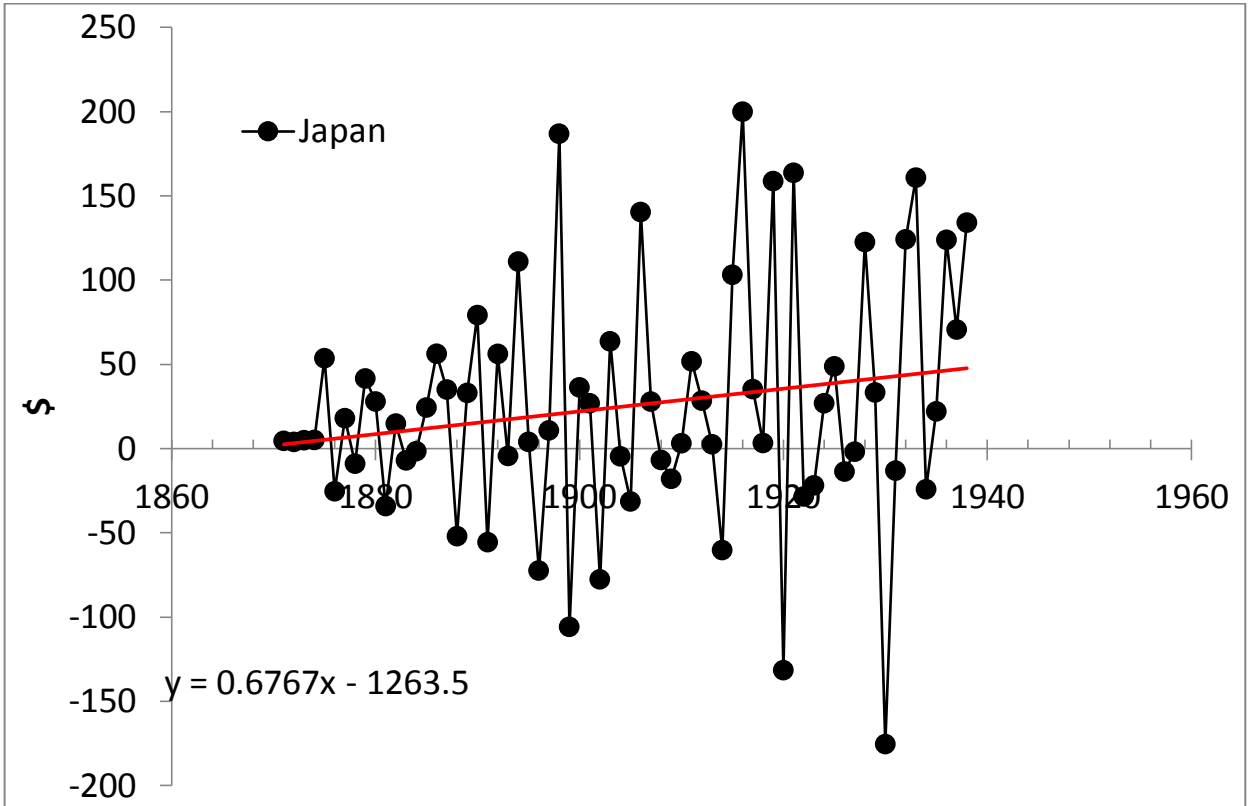

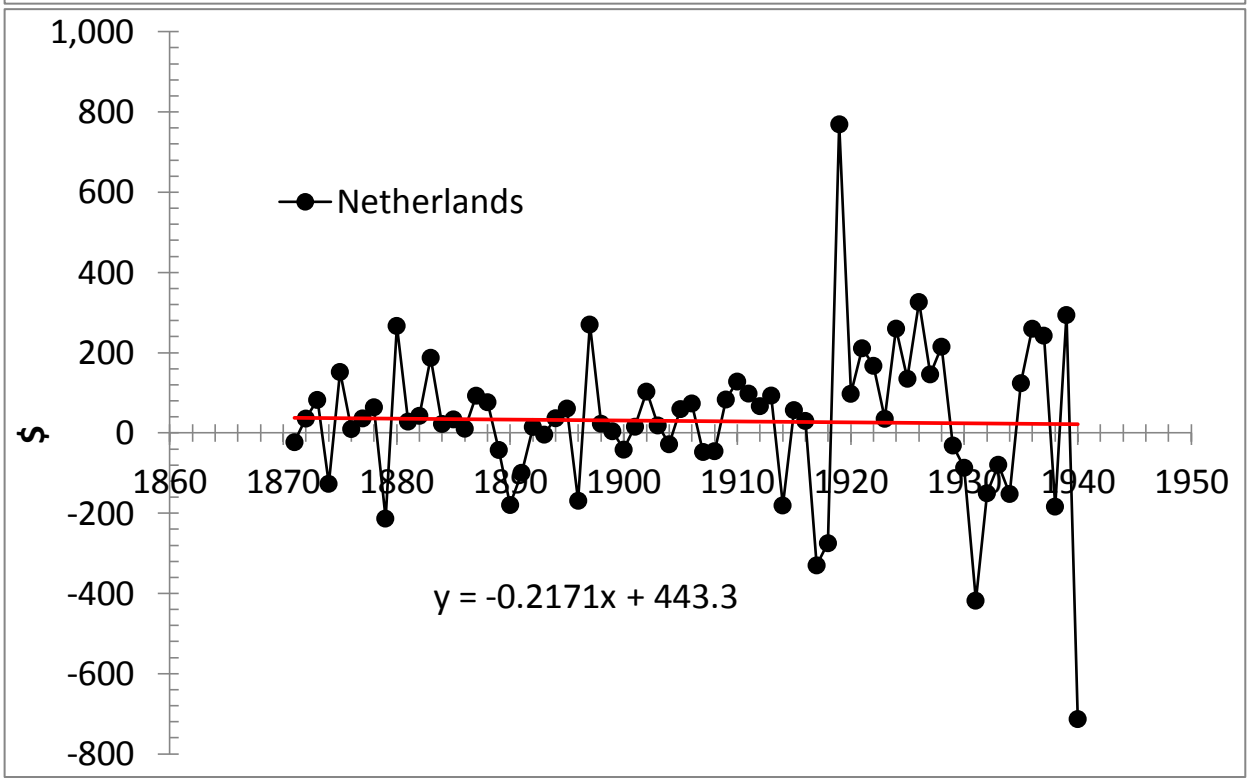

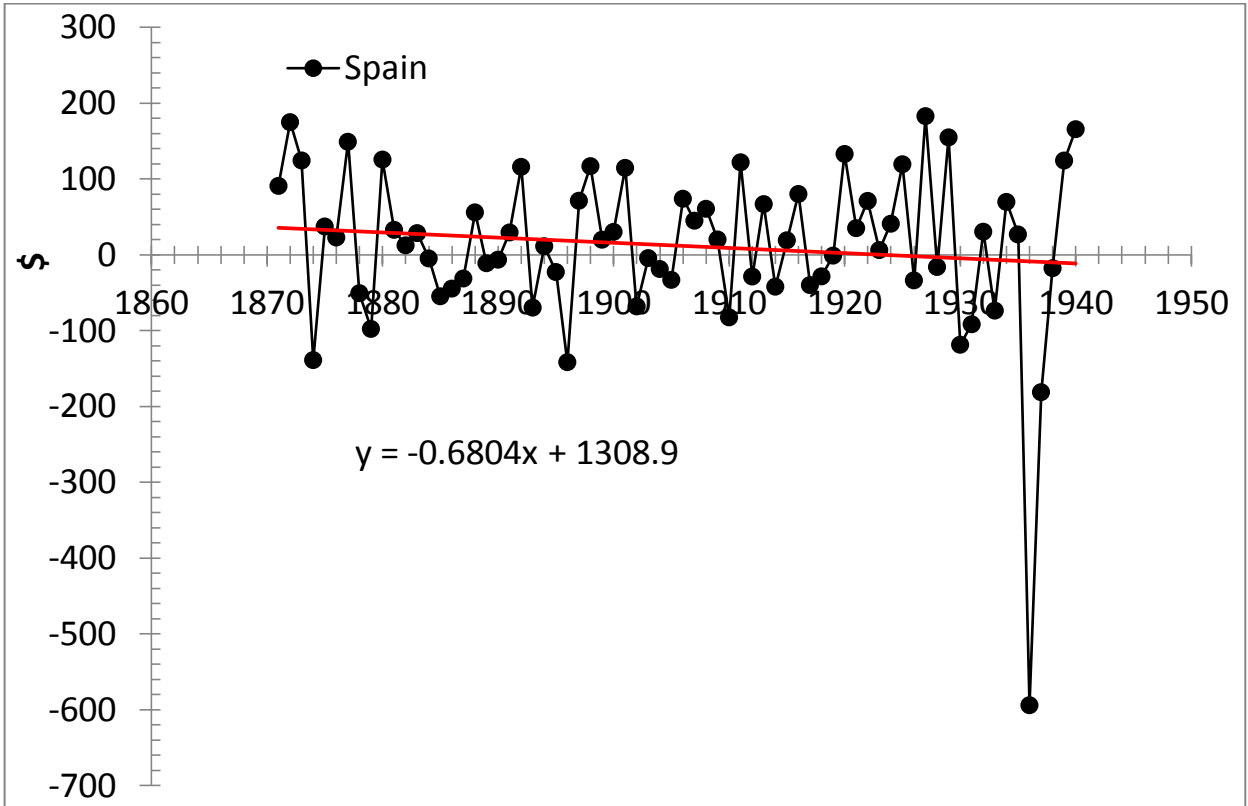
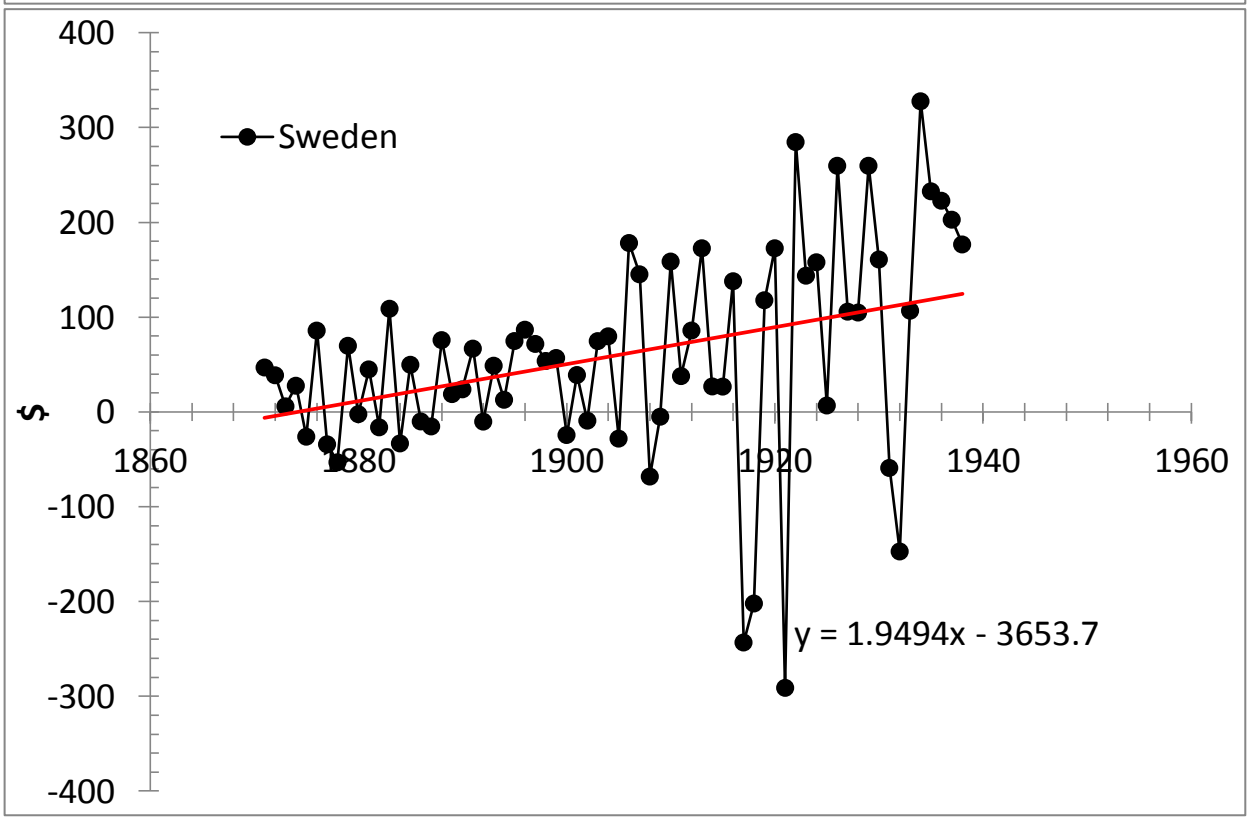

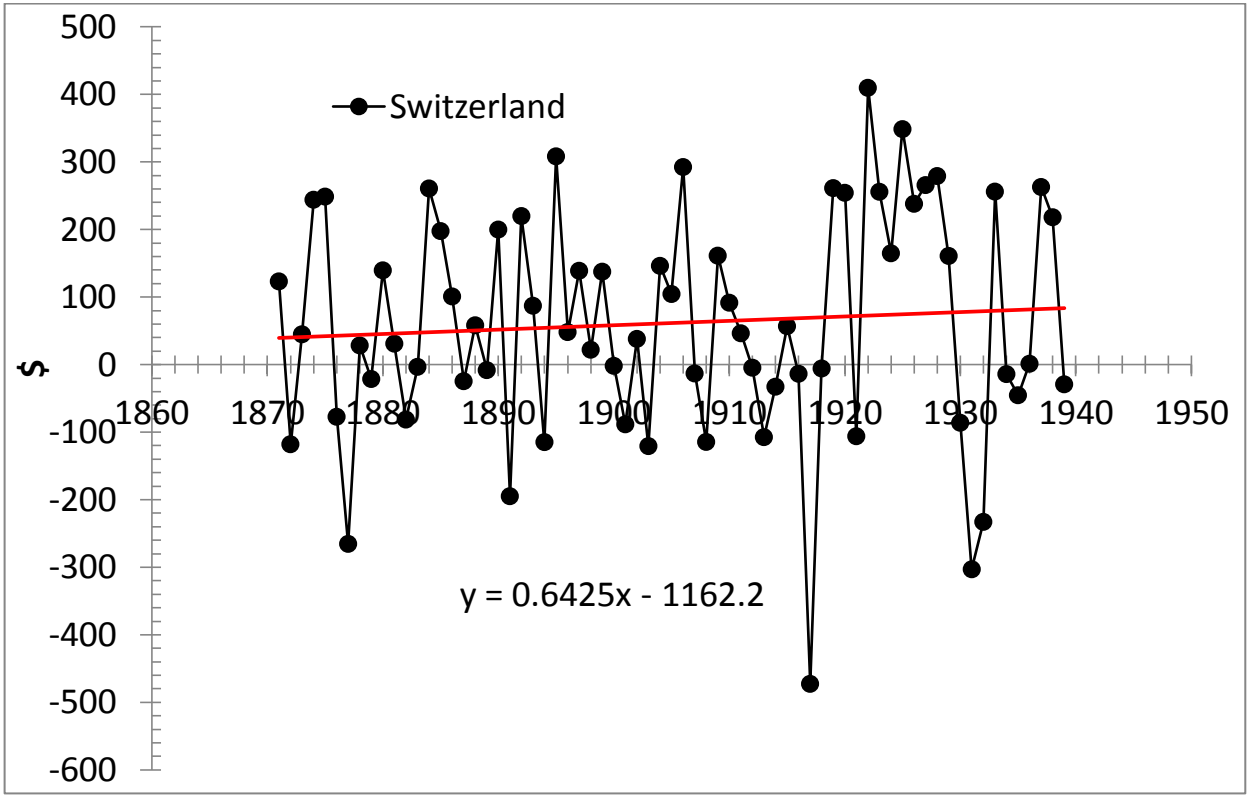
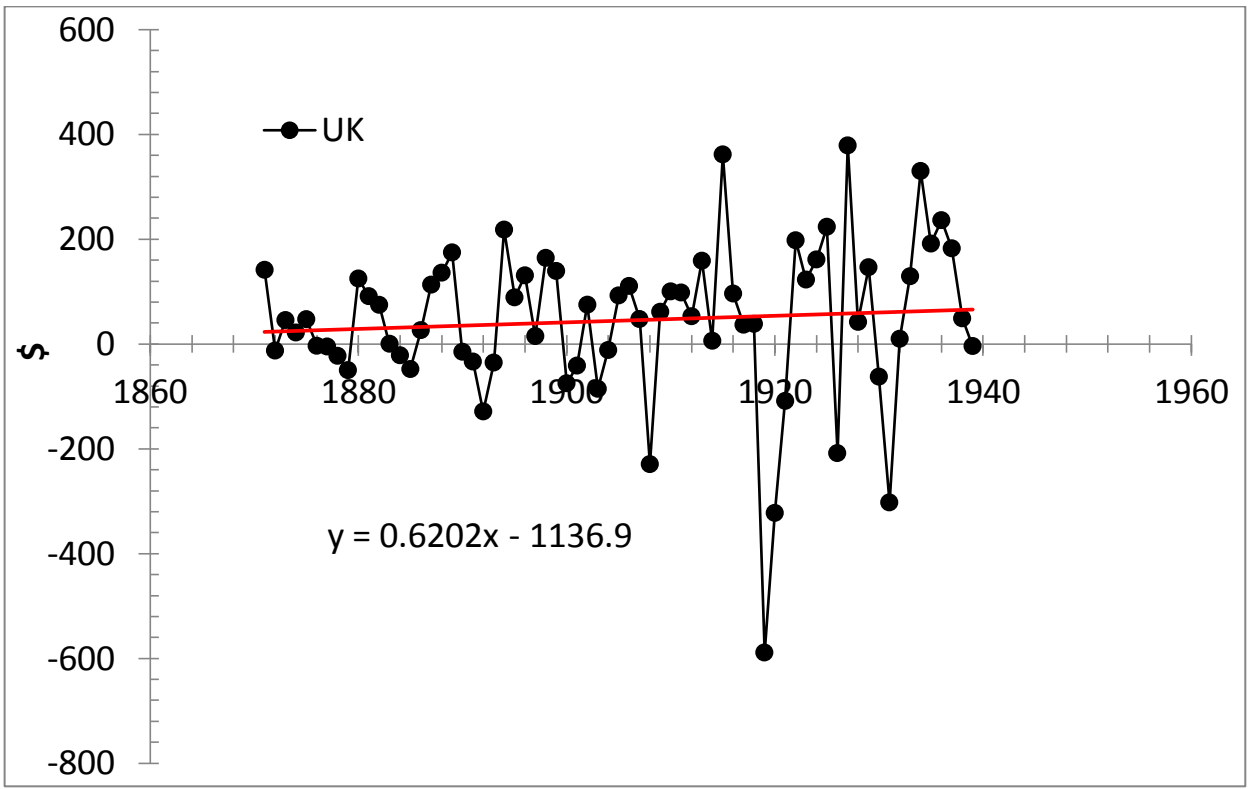

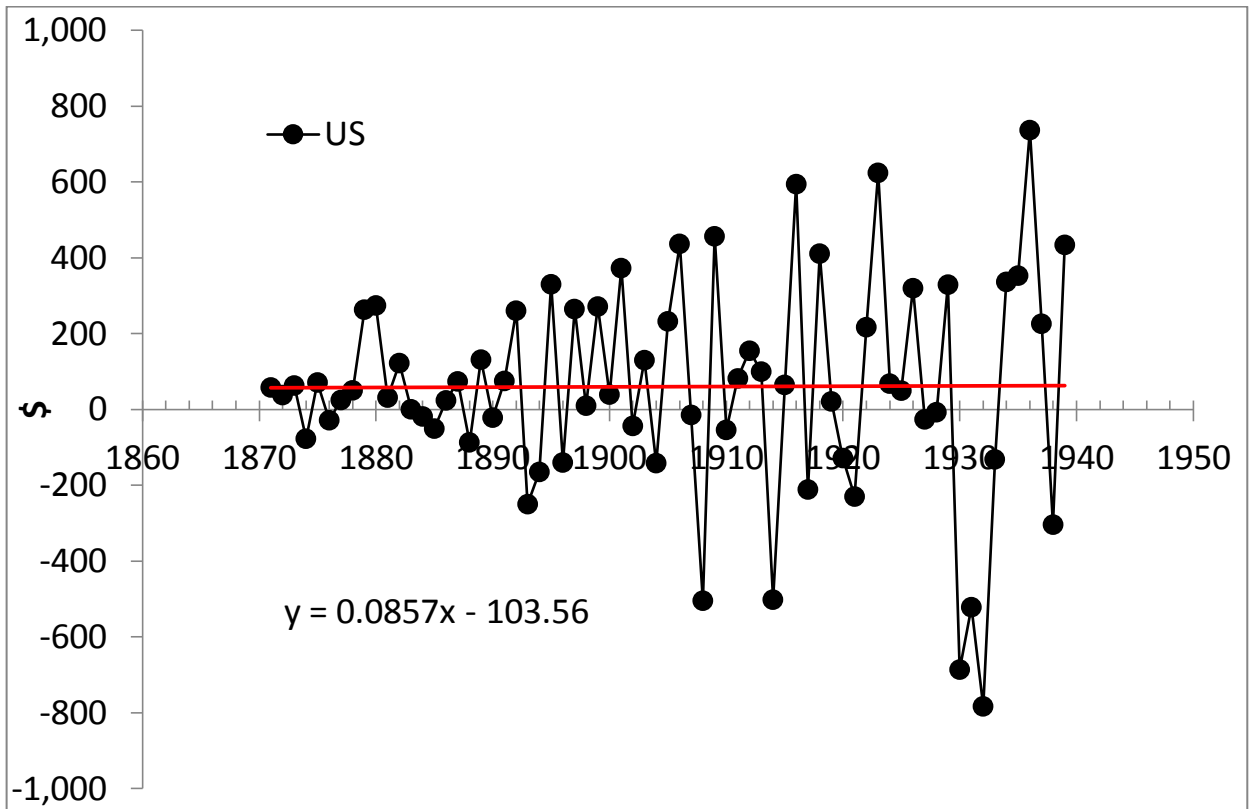
**Figure 17. Annual increment of real GDP per capita in developed countries between 1870 and 1940.**

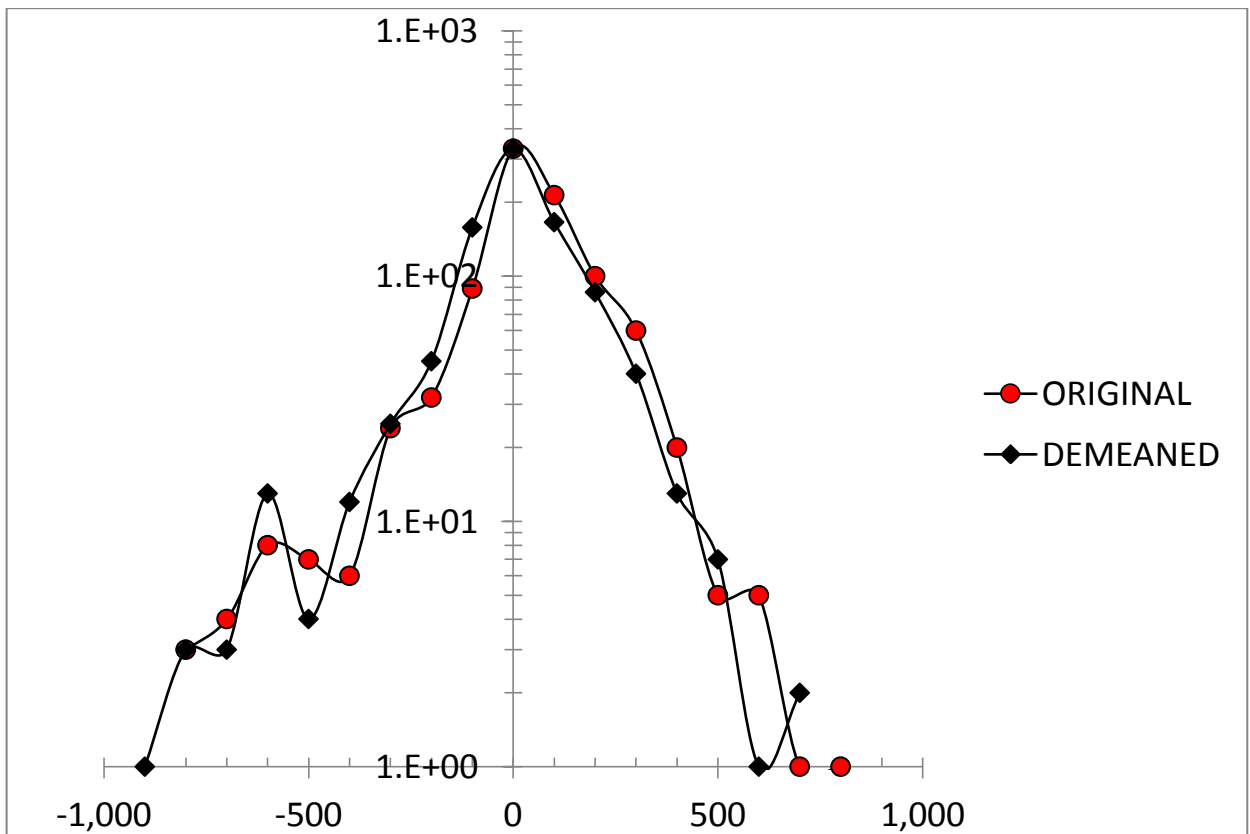
**Figure 18. Frequency distribution of annual increments from 1870 to 1940: original and demeaned**

**Conclusion**

We have addressed the most important question of real economic growth as related to the long term behaviour of real GDP per capita. The expectation of exponential growth is not supported by the whole set of observations and estimates as presented by the measured real GDP per capita since 1950 and the reconstructed time series before 1940. In both time intervals, the biggest developed economies have annual increments with small positive and negative slopes statistically not distinguishable from zero (except the case of Australia after 1950). This implies that the growth in real GDP per capita is linear in the long run. Obviously, one can formally approximate the measured time series with an exponential function with a very small exponent as related to the slope in the regression of annual increments. In no case, this would be a proof of exponential growth.

The linearity of long term growth is a significant result which is accompanied by a severe break in all time series between 1940 and 1950: the slopes and mean increments differ by an order of magnitude, although their standard deviations are close. The nature of this break should be considered in more detail, but the estimates before 1940 and after 1950 were made by different rules and using quite different methodologies: there was no working notation of GDP before 1940 and thus there was no relevant data collected systematically.

The exponential distribution of the model residuals before 1940 suggests the artificial character of the reconstructed data and also makes it more difficult to compare the periods statistically – standard statistical tests imply normal distribution of residuals and thus are biased when this assumption does not hold.